\documentstyle[aps,graphicx,epsfig,times,here]{revtex}
\begin{document}

\draft \title{Event-by-Event Fluctuations in Particle
              Multiplicities and Transverse Energy Produced
              in 158$\cdot A$ GeV Pb+Pb collisions}

\author{ 
 M.M.~Aggarwal,$^{4}$
 Z.~Ahammed,$^{2}$
 A.L.S.~Angelis,$^{7}$
 V.~Antonenko,$^{13}$
 V.~Arefiev,$^{6}$
 V.~Astakhov,$^{6}$
 V.~Avdeitchikov,$^{6}$
 T.C.~Awes,$^{16}$
 P.V.K.S.~Baba,$^{10}$
 S.K.~Badyal,$^{10}$
% ...
}

\author{(WA98 Collaboration)}

% ...
\address{$^{2}$~Variable Energy Cyclotron Centre, Calcutta
   700064, India}
% ...
\address{$^{4}$~University of Panjab, Chandigarh 160014, India}
% ...
\address{$^{6}$~Joint Institute for Nuclear Research,
   RU-141980 Dubna, Russia}
\address{$^{7}$~University of Geneva, CH-1211 Geneva
   4,Switzerland}
% ...
\address{$^{10}$~University of Jammu, Jammu 180001, India}
% ...
\address{$^{13}$~RRC ``Kurchatov Institute'',
   RU-123182 Moscow}
% ...
\address{$^{16}$~Oak Ridge National
   Laboratory, Oak Ridge, Tennessee 37831-6372, USA}
% ...

\date{\today} \maketitle

\begin{abstract}

     Event-by-event fluctuations in the multiplicities of
     charged particles and  photons, and the total transverse
     energy in 158$\cdot A$ GeV Pb+Pb collisions are studied for
     a wide range of centralities. 
     For narrow centrality bins the multiplicity and transverse 
     energy distributions are found to be near perfect Gaussians.
     The effect of detector acceptance on the multiplicity 
     fluctuations has been studied and demonstrated to follow 
     statistical considerations. The centrality dependence of the 
     charged particle multiplicity fluctuations in the measured 
     data has been found to agree reasonably well with those obtained 
     from a participant model. However, for photons the multiplicity
     fluctuations have been found to be lower compared to those
     obtained from a participant model. The multiplicity and
     transverse energy fluctuations have also been compared to those 
     obtained from the VENUS event generator.

\end{abstract}

\twocolumn


\section{INTRODUCTION}

     Fluctuations in physical observables in heavy ion collisions
     have been a topic of interest for some years as they 
     may provide important signals regarding the formation of 
     Quark-Gluon Plasma (QGP) and help to address the question of
     thermalization~\cite{fluc_gen}. With the large 
     number of particles produced in heavy ion collisions at SPS and RHIC 
     energies~\cite{qm99,qm01}, it has now become feasible 
     to study fluctuations on an event-by-event basis. Recently,
     several new methods have been proposed for the study of
     event-by-event fluctuations in various global observables
     to probe the nature of the QCD
     phase transition \cite{step,heisel,asakawa,jeon}. In a 
     thermodynamical picture of a strongly interacting system formed 
     in the collision, the fluctuations in particle multiplicities, 
     mean transverse momenta ($\langle p_{\mathrm T}\rangle$), and other
     global observables, are related to the fundamental 
     properties of the system, such as the specific heat, 
     chemical potential, and 
     matter compressibility. These, in turn, lead towards understanding the 
     critical fluctuations at the QCD phase boundary. The existence of 
     a tri-critical point at the QCD phase transition~\cite{step},  which has 
     lately been a topic of intense discussion, has been predicted to be 
     associated with large event-by-event fluctuations in the above 
     observables.

     In a first order phase transition scenario, it is believed that 
     supercooling might lead to density fluctuations resulting in
     droplet formation and hot spots \cite{droplets}. 
     These might lead to rapidity fluctuations in the form of spikes
     and gaps in the rapidity distribution. The study of event-by-event 
     fluctuations in the number of photons to 
     charged particles has also been proposed as a means to search
     for production of Disoriented Chiral Condensates (DCC)~\cite{dcc,WA98-12}.

     In nucleus - nucleus collisions the transverse energy,
     $E_{\mathrm T}$, is an extensive global variable
     \cite{wa80_et,helios,na49_et} which provides a direct measure of the 
     violence of an interaction. $E_{\mathrm T}$ is produced 
     by redirection of the longitudinal energy into transverse
     motion through interactions in which 
     the interacting particles undergo multiple scatterings and approach 
     thermalization. 
     $E_{\mathrm T}$ is also an indicator of the energy density 
     achieved in the collision. Since the energy density
     is directly related to the QGP phase transition,
     it is extremely important to study $E_{\mathrm T}$ and 
     fluctuations in $E_{\mathrm T}$. 
     Moreover it is interesting to compare the 
     fluctuations of $E_{\mathrm T}$ to those observed in the particle
     multiplicities.

     Much theoretical interest has been directed toward the subject 
     of event-by-event fluctuations, motivated by the near perfect 
     Gaussian distributions of $\langle p_{\mathrm T}\rangle$
     and particle ratios \cite{na49} measured at the SPS. 
     For these Gaussian distributions, the variance or the width of 
     the distributions contain information
     about the reaction mechanism as well as the nuclear geometry 
     \cite{step,baym,na34,wa80}. 

      The relative fluctuation ($\omega_{\mathrm X}$) in an observable $X$ 
      can be expressed as:
\begin{equation}
     \omega_X = \frac{\sigma_X^2}{\langle X \rangle},
\end{equation}
     where $\sigma_X^2$ is the variance of the distribution 
     and $\langle X \rangle$ denotes the mean value.

     The value of $\omega_{\mathrm X}$ which can be extracted from experimental
     data has contributions which originate both from trivial statistical 
     effects as well as dynamical sources.
     To extract the dynamical part associated with new physics 
     from the observed fluctuations, one has to understand the 
     contributions from statistical and other known sources. 
     Examples of known sources of fluctuations contributing to the 
     observed experimental value of $\omega_X$ include
     finite particle multiplicity, effect of limited acceptance of the 
     detectors, impact parameter fluctuations,
     fluctuations in the number of primary collisions,
     effects of re-scattering of secondaries, resonance decays,
     and Bose-Einstein correlations.
     These sources of fluctuations, 
     along with estimates of the $\omega_X$ contributions for each
     have been discussed by Stephanov et al. \cite{step} and by
     Heiselberg et al \cite{heisel}. 

     In nucleus-nucleus $(AA)$ collisions relative fluctuations in global 
     observables have been found to be smaller compared 
     to those in $pp$ collisions. It is suggested that thermal 
     equilibration in $AA$ collisions makes the fluctuations small. 
     However, the origin of fluctuations and hence the physical 
     information content are quite different in $pp$ and $AA$ collisions.
     While in $pp$ collisions one hopes to extract quantum mechanical 
     information about the initial 
     state from the event-by-event fluctuations in the final state,
     in heavy-ion collisions equilibration makes it difficult to
     achieve this goal, instead the basic aim here has been to relate the
     event-by-event fluctuations of the final state with the
     thermodynamic properties at freeze-out.

     In this article we present fluctuations in the multiplicities of
     both charged particles and photons, and in the total 
     transverse energy, over a large
     range of centralities as measured in the WA98 experiment at the CERN SPS.
     A major interest has been to
     search for fluctuations which have a new physics origin, such as
     those associated with QCD phase transition or from the formation of a DCC.

     We compare the fluctuations observed in the experimental data
     for varying centrality conditions and
     rapidity intervals to those obtained from different models.
     In the next section the WA98 experimental set up is described.
     In section~3 the criteria for the 
     centrality selection appropriate for fluctuation studies are discussed.
     Multiplicity fluctuations of photons
     and charged particles and the effect of acceptance are presented in
     section~4. In section 5, we estimate the multiplicity fluctuations in a
     participant model and compare to those obtained from data.
     Section~6 deals with transverse energy fluctuations.
     A final discussion and summary is presented in section~7.

\section{EXPERIMENT AND DATA ANALYSIS}

    In the WA98 experiment at CERN~\cite{wa98prop}, the main emphasis
    has been on high
    precision and simultaneous detection of both hadrons and photons.
    The experimental setup consisted of large acceptance hadron and photon 
    spectrometers, detectors for charged particle and photon multiplicity 
    measurements, and calorimeters for transverse and forward energy 
    measurements. The experiment took data with 158$\cdot A$ GeV Pb
    beams from the  CERN SPS in 1994, 1995, and 1996. The results
    presented here are from the Pb run in 1996 taken with the
    magnet (Goliath) turned off. The analysis makes use of
    the data taken with the photon multiplicity detector (PMD), the
    silicon pad multiplicity detector (SPMD), the mid-rapidity
    calorimeter (MIRAC), and the zero degree calorimeter (ZDC). 

    The circular Silicon Pad Multiplicity Detector (SPMD), used for 
    measuring charged particle multiplicity, was located 32.8 cm from
    the target. It had full azimuthal coverage in the region 
    $2.35 \le \eta \le 3.75$.
    The detector had four overlapping quadrants, each fabricated
    from a single 300~{$\mu m$} thick silicon wafer.
    The active area of each quadrant was divided into 1012 pads forming 
    46 azimuthal wedges and 22 radial bins with pad size increasing with 
    radius to provide a uniform pseudo-rapidity coverage. The
    intrinsic efficiency of the detector was better than $99~\%$.
    During the data taking, $95~\%$ of the pads worked properly.
    It was nearly transparent to
    high energy photons, since only about $0.2~\%$ are 
    expected to interact in the silicon. 
    Details of the characteristics of the SPMD
    can be found in Ref. \cite{WA98-3,spmd_nim}.

    The photon multiplicity was measured using the preshower photon 
    multiplicity detector (PMD) placed at a distance of 21.5 meters
    from the target. The detector consisted of 3 radiation length ($X_0$)
    thick lead converter plates placed in front of an array of square 
    scintillator pads of four different sizes, varying from 
    15 mm$\times$15 mm to 25 mm$\times$25 mm, placed in 28 box modules.
    Each box module had a matrix of $38\times$50 pads which were 
    read out using one image intensifier + CCD camera system. 
    Details of the design and characteristics of the PMD may be found 
    in Ref. \cite{pmd_nim,WA98-9}. The results presented here make use 
    of the data from the central 22 box modules covering the
    pseudo-rapidity range $2.9\le \eta\le 4.2$. The clusters of hit pads,
    having total ADC content above a hadron rejection threshold were 
    identified as photon-like. Detailed simulations showed that the
    photon counting efficiencies for the central to peripheral cases 
    varied from $68\%$ to $73\%$. The purity of the photon sample,
    $N_{\gamma-{\mathrm like}}$, in the two cases varied from $65\%$ to $54\%$.

    The transverse energy was measured with the MIRAC 
    calorimeter \cite{awes} placed at 
    24.7 meters downstream from the target. It consisted of 30 stacks, each
    divided vertically into 
    6 towers, each of size 20 $\times$ 20 cm$^2$, and segmented 
    longitudinally into an electro-magnetic (EM) and a hadronic section. 
    The depth of an EM section was 15.6$X_0$ (equivalent to $51\%$ of 
    an interaction length)
    which ensured almost complete containment of the
    electromagnetic energy ($97.4\%$ and $91.0\%$ containment
    calculated for 1~GeV and 30~GeV photons, respectively). 
    The MIRAC was used to measure both the transverse 
    electromagnetic ($E_{\mathrm T}^{em}$) 
    and hadronic ($E_{\mathrm T}^{had}$) energies in the 
    interval $3.5\le\eta\le 5.5$ with  
    a resolution of $17.9\%/\sqrt E$ and $46.1\%/ \sqrt E$,
    ($E$ in GeV), respectively. The $E_{\mathrm T}$ provides a measure of the 
    centrality of the reaction. 
    Events with large $E_{\mathrm T}$ correspond to very central
    reactions with small impact parameter and vice versa.

    The Zero Degree Calorimeter (ZDC) measured 
    the total forward energy, $E_{\mathrm F}$, at
    $\theta\le 0.3^{o}$ with a resolution of 80\%/$\sqrt E + 1.5\%$, with 
    $E$ expressed in GeV. $E_{\mathrm F}$ provides complementary 
    information on the centrality with low $E_{\mathrm F}$ energy
    deposit corresponding to small impact parameter collisions.

    For the results to be presented below, the following sources
    of errors have been included in the systematic
    error estimates: \\

\noindent {\it {Errors in $N_{\gamma-{\mathrm like}}$ }:} \\

\noindent {(a) The major source of error in $N_{\gamma-{\mathrm
      like}}$ is due to the effect of clustering of the pad signals.
      This error is determined from the simulation by comparing the
    number of known tracks on the PMD with the total number of 
    photon-like clusters. 
    The  result is that
    the number of clusters exceeds the number of tracks by
    3\% in the case of
    peripheral events and by 7\% for high multiplicity central events
    ~\cite{WA98-9}.} \\

\noindent {(b) The uncertainty in the ADC value of the hadron rejection
      threshold in the PMD leads to an error in the estimation of
      $N_{\gamma-{\mathrm like}}$ clusters. The hadron rejection
      threshold has been set at three times the MIP (minimum ionizing
      particle) peak. The value of MIP peak was changed by 10\% of the
      peak value (3 ADC) in order to estimate the
      systematic error. The error in $N_{\gamma-{\mathrm like}}$
      value is 2.5\% \cite{WA98-9}.} \\

\noindent {(c) The error because of the variation in scintillator
       pad-to-pad gains is found to be less than 1\%.} \\

   The combined systematic error on $N_{\gamma-{\mathrm like}}$ is
   asymmetric and centrality dependent. The errors are 
   -3.2$\%$ and +3.4$\%$ for peripheral collisions and
   -7.1$\%$ and +3.0$\%$ for central collisions.
   The errors on $N_{\gamma}$, obtained after correcting 
   for photon counting efficiency and purity of photon-like sample, will be 
   discussed in section~V. \\

\noindent {\it {Errors in $N_{\mathrm ch}$ } :} \\

The uncertainty in the  $N_{\mathrm ch}$ obtained from SPMD 
has been discussed in detail in Ref.~\cite{WA98-3}. The total error  
has been estimated to be about 4\%. \\

\noindent { \it {Errors in centrality selection through $E_{\mathrm
      T}$ and $E_{\mathrm F}$ } :} \\

    The centrality of the interaction is determined by the total 
    transverse energy ($E_{\mathrm T}$) measured in the MIRAC.
    The finite resolution in the measurement of $E_{\mathrm T}$
    contributes to the systematic error. 
    For the analysis of fluctuation in $E_{\mathrm T}$ which uses
    MIRAC data directly, the centrality is determined by the forward
    energy, $E_{\mathrm F}$.
    The finite resolution in the measurement $E_{\mathrm F}$
    contributes to the systematic error in $E_{\mathrm T}$ fluctuations.
    These errors are centrality dependent. \\

\noindent {\it {Fitting errors } :} \\

      The fitting errors associated with the determination of the
      fit parameters  of the multiplicity and transverse energy 
      distributions also contribute to the 
      final systematic error in both the photons and charged particles 
      and transverse energy respectively. The maximum contribution of
      this error to the fluctuation was found to be $2\%$.

\section{CENTRALITY SELECTION FOR FLUCTUATION STUDIES}

    The centrality of the interaction was determined by the total 
    transverse energy measured in the MIRAC. For the part of the analysis 
    where transverse energy data are used for the fluctuation studies, 
    the centrality was determined instead by the 
    forward energy, $E_{\mathrm F}$, as measured in the ZDC. 
    The centralities are expressed as fractions of the minimum bias 
    cross section as a function of the measured total 
    transverse energy using MIRAC, or total forward energy using the ZDC. 
    Fig. \ref{et_zdc}(a) and (b) show the 
    minimum bias distributions of $E_{\mathrm T}$ and $E_{\mathrm F}$,
    respectively. The arrows in 
    the figures indicate the values of $E_{\mathrm T}$ and $E_{\mathrm F}$
    for the top (most central) $1\%$,  $2\%$, $5\%$, 
    and $10\%$ of the minimum bias cross section.
    Predictions from VENUS 4.12~\cite{venus} are also shown as solid
    histograms. This will be discussed in a later section.

    The anti-correlation of $E_{\mathrm T}$ and $E_{\mathrm F}$ 
    is shown in \ref{et_zdc}(c). It illustrates
    that either $E_{\mathrm T}$ or $E_{\mathrm F}$ can be used 
    nearly equivalently to define the centrality of the reaction.

    Fig. \ref{nch_ngam}(a) and \ref{nch_ngam}(b) 
    show the minimum bias distributions for $\gamma-{\mathrm like}$ 
    clusters and charged particles, respectively, for the full
    acceptances of the two detectors (PMD and SPMD). The multiplicity
    distributions
    corresponding to the centrality cuts using the total $E_{\mathrm T}$ for
    the top $1\%$, $2\%$, and $5\%$ of the minimum bias cross section are 
    also shown in the same figures.
    These distributions have been fitted to Gaussians. The extracted fit
    parameters are used for the analysis of the fluctuations.

  Fig.~\ref{cen_nch_ngam} shows the variation of the mean ($\mu$), 
  standard deviation ($\sigma$), and chi-square per degree of freedom 
  (${\chi^{2}/ndf}$) of the photon and charged particle multiplicity
  distributions for different centrality bins.
  Here the centrality class is chosen with increasing width, as $0-1\%$, 
  $0-2\%$, $0-3\%$,$\cdots$,$0-10\%$.
  As expected, the mean value decreases
  and the sigma increases as we make broader centrality selection to include
  more of the cross section.
  From the ${\chi^{2}/ndf}$ values, one observes that the
  distributions increasingly deviate from Gaussians with increasing width in
  the centrality selection. 
  For a centrality selection width of greater than $5\%$, 
  the ${\chi^{2}/ndf}$ rises above $2$.
  The variation of $\mu$ and $\sigma$ indicate that 
  the extracted relative fluctuation ($\omega_{X}$)
  will grow with the increase in the width of the centrality 
  selection interval.
  This indicates that the impact parameter
  fluctuations will dominate as the centrality selection is broadened.
  From this we conclude that the centrality selections should be made 
  with as narrow bins in $E_{\rm T}$ as possible, such that the
  multiplicity  distributions
  are good Gaussians and the impact parameter fluctuations are minimized.
  With this in mind we have used centrality selection bins 
  of $2\%$ widths in cross section, taken as, $0-2\%$,
  $2-4\%$, $\cdots$, $62-64\%$.

  Fig.~\ref{cen2_nch_ngam} shows the variation of  $\mu$, $\sigma$, 
  and ${\chi^{2}/ndf}$ of the photon and charged particle multiplicity 
  distributions  within 
  the full acceptance of the detectors with these narrow bins in centrality. 
  The data presented here cover
  the region from central (top 2\% of the minimum bias
  cross sections) to peripheral collisions (up to 65\% of the minimum
  bias cross section where the average number of participants is $26$).  
  It is seen that both the $\mu$ and $\sigma$
  values decrease  towards peripheral collisions. 
  The ${\chi^{2}/ndf}$ values are mostly in the region between
  1.0 and 2.0  over the entire range of centralities  
  considered. This suggests that narrow cross section slices in the 
  $E_{\mathrm T}$ or $E_{\mathrm F}$ distributions are necessary
  to study the multiplicity fluctuations and minimize the
  influence from impact parameter fluctuations.

\section{MULTIPLICITY FLUCTUATIONS AND THE EFFECT OF ACCEPTANCE}

     The relative fluctuations in multiplicity 
     for $\gamma-{\mathrm like}$ clusters
     and charged particles have been calculated using the 
     $\mu$ and $\sigma$ values
     from Fig.~\ref{cen2_nch_ngam} and Eq.~1. These values are shown
     in Fig.~\ref{2perc_W} 
     as functions of centrality, for full coverage of
     PMD ($2.9 \le \eta \le 4.2$) and SPMD ($ 2.35 \le
     \eta \le 3.75$), respectively.
     The errors shown in
     the figures are systematic errors, the sources of which has been 
     already discussed in previous section.
     For both $\gamma-{\mathrm like}$ clusters and charged particles
     the relative fluctuations are seen to increase in going 
     from central to peripheral
     collisions. However, for charged particles the increase is much
     stronger.

    In order to make a direct comparison of the fluctuations of 
    photons and charged particles, the multiplicities should be
    studied in the region of common coverage of the detectors 
    in terms of both pseudo-rapidity and azimuthal angle. In a 
    later section we will compare the results obtained from data
    with those from model calculations for the common coverage.
    The region of common coverage of the two detectors in the WA98 
    experiment was  $0.85$ units in $\eta$ ($2.9\le \eta \le 3.75 $).
    The general trend of the variation of the Gaussian fit parameters,
    with the reduced number of particles, for the
    common coverage is found to be similar to that 
    obtained with full coverage for each  detector.
    Fit parameters $\mu$, $\sigma$, and ${\chi^{2}/ndf}$, as obtained
    for centrality bins of  $2\%$ in width, for the common coverage of the two
    detectors are shown in Fig.~\ref{com_ngam_nch}. 
    As was the case for the larger acceptance, the $\mu$ and $\sigma$ 
    values are seen to decrease towards more peripheral 
    event selection. The ${\chi^{2}/ndf}$ values are reasonable. 
    Using the above values of the Gaussian parameters together with 
    Eq.~1, the relative multiplicity
    fluctuations were calculated and are shown in Fig.~\ref{com_fluc}.
    The error shown include the fit errors as well as the other systematic
    errors discussed earlier.
    The relative fluctuations for 
    $\gamma-{\mathrm like}$ clusters is seen to be rather constant 
    over the full centrality
    range with an average value of $2.2\pm 0.21$.
    In comparison, the relative fluctuations for charged particles is
    $1.56 \pm 0.13$ at the most 
    central bin ($0-2\%$) increasing to $2.8 \pm 0.16$ for 
    the least central bin (62-64\%).

  Following this discussion of 
  the fluctuations in the multiplicity of photons and 
  charged particles for
  the full acceptance regions and for the regions of 
  common coverage of the photon and
  charged particle detectors, we now analyze the effect
  of detector acceptance on the observed fluctuations in more detail. 
  For this we have taken two
  different $\eta$ coverage regions for each detector.
  For the PMD the $\eta$ ranges chosen are
  $ 3.0 \le \eta \le 4.0$ and $ 3.25 \le \eta \le 3.75$ 
  (with full $\phi$ coverage). 
  The resulting $\omega$ values for the two cases are 
  shown in  Fig.~\ref{fluc_acc}(a).
  Qualitatively, the variation of the fluctuations with centrality
  is similar for both coverages, but the magnitude of the relative 
  fluctuations is lower for smaller $\eta$ coverage.

  For the SPMD, the fluctuations were calculated for the rapidity intervals
  $ 2.35 \le \eta \le 3.35$ and  $ 2.65 \le \eta \le
  3.15$. These bins have width of one and one half unit 
  in $\eta$ around mid-rapidity.
  The results are shown in
  Fig.~\ref{fluc_acc}(b). Qualitatively, the results are again
  similar to each other with the magnitude of the relative
  fluctuations decreasing as the coverage
  in $\eta$ is decreased.

  The decrease in the relative fluctuations as the acceptance 
  is decreased can be understood
  in terms of a simple statistical picture \cite{acp_fluc}. 
  Assume that there are $m$ particles
  produced in the collision out of which $n$ particles are accepted 
  randomly into the detector acceptance. 
  In this case, the distribution of $n$, will follow a binomial distribution 
  with mean $mf$ and variance $mf(1-f)$ where $f$ is the
  fraction of particles accepted.  
  Therefore the fluctuations in the number of particles 
  accepted for a fixed ($m$) number of particles produced is $(1-f)$.
  In principle $m$ can have an arbitrary distribution as given by $P(m)$ with
  known first and second moments. 
  The fluctuations in the $n$ accepted particles out of the $m$
  particles produced is then given as:
\begin{equation}
\omega_n = 1 - f + f\omega_m
\end{equation}

   Thus considering the fluctuations in one unit of $\eta$ as $\omega_m$ 
   we can calculate the expected fluctuations for one half unit of $\eta$
   using the above equation. Here $f$ corresponds to the ratio
   of the total number of particles accepted in one half unit of $\eta$
   coverage to that accepted over one unit in $\eta$. 
   For the acceptance regions used, the 
   average value of $f$ for photons is about $0.52$ and that of
   charged particles is about $0.54$. Using these values in Eq.~2
   we can calculate the expected fluctuations for half unit of $\eta$ coverage
   from the results for one unit of $\eta$ coverage, under the
   assumption of a binomial sampling. As shown in 
   Fig.~\ref{fluc_acc} the empirical calculations almost exactly reproduce
   the observed result in the narrower acceptance window for charged
   particles and agrees reasonably well within the quoted errors for photons.

\section{ESTIMATION OF FLUCTUATIONS IN A PARTICIPANT MODEL}

     In a picture where the nucleus-nucleus collision is thought of as the sum
     of contributions from many sources  
     created in the early stage of the interaction, 
     the variance of the distribution of any observable has 
     contributions from:

\begin{itemize}

   \item   the fluctuations in the number of sources, largely due 
           to different impact parameters. Even if the 
           impact parameter window is narrowed, density fluctuations within 
           the nucleus will make this contribution non-zero. 

    \item  the fluctuations in the number of 
           particles produced by each source.  
           Quantum fluctuations in the $NN$ cross section can lead
             to such effects. 

    \item  the fluctuations due to any dynamical process or 
           critical behavior in the evolution of the  system.

\end{itemize}

     The contribution from the first two effects
     leads to fluctuations in the number of participant nucleons which may be
     related to the initial size of the interacting system before it 
     thermalizes.
     Resonance decays have also been shown to increase the multiplicity
     fluctuations by a large factor \cite{step,heisel}.

     Following a simple participant model \cite{heisel,baym,na34,hwa,van}, 
     the particle multiplicity (of photons or charged particles), $N$,
     may be expressed as :
\begin{equation}
        N =  \sum_{i=1}^{N_{\mathrm part}}  n_i
\end{equation}
  where ${N_{\mathrm part}}$ is the number of participants 
  and $n_i$ is the number of
  particles produced in the detector acceptance by the $i^{th}$ participant.
  On an average, the mean value of $n_i$ is the ratio of the 
  average multiplicity
  in the detector coverage to the average number of participants, i.e.,
  $\langle n \rangle = \langle N \rangle / \langle N_{\mathrm part} \rangle $.

  Thus the fluctuations in $N$ will have contributions due to fluctuations
  in $N_{\mathrm part}$ ($\omega_{N_{\mathrm part}}$) and also due to the
  fluctuations in the number particles produced per participant
  ($\omega_{n}$). Again, the fluctuations of $n$ given 
  as $\omega_{n}$, 
  will have a strong dependence on the acceptance of the       
  detector. In the absence of correlations between  the $n_i$'s, the
  multiplicity fluctuations, $\omega_{N}$,
  can be expressed as

\begin{equation}
       \omega_{N}  =  \omega_{n} + \langle n \rangle \omega_{N_{\mathrm part}}
  \end{equation}

  Comparison of data with the results of such model calculations might
  reveal the extent to which 
  the principle of superposition of nucleon-nucleon 
  (NN) interactions is valid in the case of heavy ion collisions.
  The participant model is expected to hold reasonably well for
  peripheral collisions where there are only few NN collisions, while for
  central collisions the particle production gets affected by
  NN scattering, rescatterings between produced particles, energy degradation, 
  and other effects. Next we discuss the calculation of each of the terms in 
  Eq.~4. 

\subsection{\it Calculation of $\omega_{N_{\mathrm part}}$ } 

 The impact parameter fluctuations are reflected in the fluctuations
 in the number of participants. We have estimated this contribution using
 the VENUS 4.12 event generator with default setting.
 A set of 100K minimum bias  Pb+Pb events at 158$\cdot A$ GeV was 
 generated for calculation of the number of participants.
 To match the centrality selection of the reaction in simulation to 
 that in data, we have carried out a fast simulation in which  
 $E_{\mathrm T}$ from VENUS was calculated within MIRAC coverage 
 taking the resolution factors for the hadronic and electromagnetic 
 energies of MIRAC into account. The corresponding $E_{\mathrm T}$ 
 distributions for VENUS are shown as the solid curve in 
 Fig.~\ref{et_zdc}(a). It is seen that the agreement with data is 
 quite reasonable.

 The distributions of $N_{part}$ for the same narrow ($2\%$) bins 
 of centrality, as discussed above for the data, are well described
 by Gaussian distributions. 
 Fig.~\ref{npart_fluc_2} shows the variation of  $\mu$, $\sigma$, 
 ${\chi^{2}/ndf}$, and relative fluctuation $\omega_{N_{\mathrm part}}$ 
 calculated from the fit parameters with the $2\%$ bins in centrality.
 One can see that the relative fluctuation in the number of participants,
 $\omega_{N_{\mathrm part}}$, is around $1$. 
 The statistical errors are small and are within the size of the 
 symbols.

 The systematic errors shown in the figures, have contribution from the 
 following sources, which have been added in quadrature.

\begin{itemize}

\item Nucleon density distribution: In order to
      estimate the error due to this we have calculated the number of
      participants from VENUS (as shown in the figure) and those from 
      FRITIOF. The difference for each centrality bin was considered 
      as representative of the error \cite{WA98-10}.

\item Finite resolution of $E_{\mathrm T}$:
      Systematic errors due to this were calculated by varying the
      centrality as per the MIRAC resolution \cite{WA98-9}.

\item Fitting errors:  Errors associated with the determination of the
      fit parameters  of the Gaussian distributions also contribute to the 
      final systematic error in the number of participants.

\end{itemize}

 The quantity $\langle n \rangle $ is equal to 
 the ratio of the mean charged particle (or photon) multiplicity 
 for a given acceptance to the mean number of
 participants for the same centrality bin. Thus the contribution from 
 the term   
 $\langle n \rangle \omega_{N_{\mathrm part}}$ to the total
 fluctuations (Eq.~4) can be easily obtained.

\subsection{\it Calculation of $\omega_{n}$ }

  This term gives the fluctuations in the number of particles produced 
  per participant. It has a strong dependence on acceptance as given
  earlier in Eq.~2 and shown in Fig.~\ref{fluc_acc}. 
  To calculate $\omega_{n}$ as per Eq.~2 we next obtain the terms $f$
  and $\omega_{m}$.
  The quantity $f$ is the ratio of the number
  of particles per participant accepted within the acceptance of the detector 
  ($\langle n \rangle$) to the total number of particles produced per 
  participant   ($\langle m \rangle$). The value
  of $\langle n \rangle$  for each centrality bin and for a given
  acceptance can be calculated as discussed in the preceding section. 
  To obtain the value of $\langle m \rangle$ we make use of  
  the data existing in the literature for nucleon-nucleon (NN)
  collisions. As discussed
  in Refs. \cite{ua5,whit}  the mean number of charged particles 
  and photons produced in nucleon-nucleon collisions can be parametrized as a 
  function of cms energies ($\sqrt{s}$ from 2~GeV to 500~GeV ) in
  the following manner :
\begin{equation}
    {\langle N_{\mathrm ch} \rangle}^{NN}  =  
             -4.7 (\pm 1.0) + 5.2 (\pm 0.8) s^{0.145 (\pm 0.01)}
\end{equation}
\begin{equation}
       {\langle N_{\mathrm \gamma} \rangle}^{NN}  =  
             -9.9 (\pm 2.1) + 8.5 (\pm 1.9) s^{0.113 (\pm 0.015)}
\end{equation}

  For the 158$\cdot A$ GeV SPS energy discussed here
  this parameterization gives the 
  average charged particle multiplicity to be $7.2$ with the corresponding
  number for photons being $6.3$. Thus the average charged particle and 
  photon multiplicities per participant are $3.6$ and $3.15$ respectively.

  In addition, $\sigma$ for the charged particle multiplicity in 
  nucleon-nucleon collisions
  shows a linear dependence with the average charged particle multiplicity as
  $0.576 (\langle N_{\mathrm ch}^{NN} \rangle  - 1)$, 
  as given in Ref.~\cite{whit}. 

  This can be used to calculate  $\omega_{m}$, which is given as:
\begin{equation}
      \omega_{m} = 0.33 \frac{(\langle N_{\mathrm ch}\rangle
        -1)^2}{\langle N_{\mathrm ch}\rangle }
\end{equation}

  For charged particles at SPS energies this gives a value of 
  $\omega_{m} = 1.8$. 
  However, for photons this number is not known since there is no
  similar parameterization.
  In the absence  of such a parameterization of $\sigma$ 
  for photons we will also assume that $\omega_{m} = 1.8$ for the
  photon multiplicity. Fluctuations
  of photons, in principle, are expected to be similar to those
  for charged particles. This is because the majority of photons 
  come from decay of $\pi^0$ and while the majority of charged particles 
  are charged pions ($\pi^{\pm}$). 

  From the values of $\langle n \rangle $, $\langle m \rangle $, 
  and $\omega_{m}$ for a given acceptance and
  centrality, the term $\omega_{n}$ can then be calculated. 

\subsection{\it Comparison of data to model calculations } 

 We first compare the experimental results of multiplicity fluctuations
 to those of the calculations using the participant model in the
 common coverage of PMD and SPMD.
 Fig.~\ref{ch_fluc_model} shows a comparison of the fluctuation
 in charged particles from data to that obtained 
 from the calculations using the model described above. The results 
 are plotted as a function of the number of participants. The 
 horizontal errors on the number of participants are shown only on 
 the data points. The error 
 on $\omega$ calculated in the model is mainly due to the error on 
 the mean number of charged particles in nucleon-nucleon interactions,
 the error in the number of participants calculated, and the uncertainty 
 in the simulation of the calculated transverse energy. 
 For clarity of presentation we have given results corresponding to
 alternate $2\%$ centrality bins, i.e $0-2\%$, $2-4\%$, $\cdots$, $62-64\%$.
 The results from VENUS are also shown in form of solid
 line in Fig.~\ref{ch_fluc_model}, and are found to remain almost
 constant over the entire centrality range.
 Charged particle fluctuations determined from data and the participant
 model decrease in going from peripheral to central collisions,
 although the dependence on centrality is weaker for the model
 calculation. 

 Fig.~\ref{gam_fluc_model} shows fluctuations in the $\gamma$-like
 clusters as well as $N_\gamma$ after the correction. 
 The results, plotted as a function of the number of participants,
 are compared to those of the participant model 
 calculations for photons and results from VENUS.
 Using the estimated values of efficiency ($\epsilon_{\gamma}$) 
 and purity ($f_p$), the number of photons in an event is calculated
 by using the relation:
\begin{equation}
      N_{\gamma} = \frac{f_p}{\epsilon_{\gamma}}~N_{\gamma-{\mathrm like}} 
\end{equation}
 The photon counting efficiency in PMD
 varies from $68\%$ to $73\%$ for central to
 peripheral collisions. The purity of the measured photon sample
 varies from $65\%$ to $54\%$ for central to peripheral collisions.

 The systematic errors associated with $N_{\gamma-{\mathrm like}}$ 
 have already been discussed in section~II. 
 The additional errors in the conversion from $N_{\gamma-{\mathrm like}}$ 
 to $N_{\gamma}$ are mainly due to  errors in 
 estimation of photon counting efficiency and purity.
 These sources of these errors are given below:
\begin{itemize}

\item Event-by-event variation in photon counting efficiency
      ($\epsilon_{\gamma}$) and purity of photon sample ($f_p$).
      These have been found to vary from $3\%$ to $6\%$ for
      central to peripheral collisions.

\item The purity factor depends on the ratio of the number of photons
    and charged particles within the PMD coverage. The
    systematic error associated with this ratio has been
    studied by using the FRITIOF \cite{fritiof} event generator in addition
    to VENUS. 
    The average photon multiplicity estimated
    by using FRITIOF is found to be higher
    by about 4\% in peripheral and by 1\% in central collisions, compared to
    the values obtained using VENUS.

\item The photon counting efficiency determined in the present case relies on
    the energy spectra of photons as given by the VENUS event generator. As
    the conversion probability for low energy photons falls sharply 
    \cite{wa93nim} with decreasing energy below 500 MeV, the estimate of
    $\epsilon_{\gamma}$ may be affected if the energy spectra  
    in the actual case is different. 
    Photon energy spectra have been measured by the WA98 lead glass
    calorimeter. By extrapolating these measured spectra to the PMD
    acceptance we have estimated the photon counting efficiencies for
    different eta bins and centralities. These results turn out to be lower
    compared to those obtained from VENUS by 2-9\% for central events and
    3-13\% for peripheral events, the smaller value corresponds
    to larger 
    pseudo-rapidity region of the PMD acceptance. The average difference
    in efficiencies 
    within the PMD acceptance are 6\% for central and 9\% for peripheral
    collisions. These differences add to
    the systematic errors on the photon counting efficiency. 
\end{itemize}
  The total systematic error on the multiplicity of photons
  ($N_\gamma$) are -6.7$\%$ and +12.5$\%$ 
  for peripheral collisions and 
  -8.0$\%$ and +9.0$\%$ for central collisions.

  The fluctuations in the number of photons have been 
  estimated from the fluctuations in the number of photon-like clusters by
  using Eq.~8:
\begin{equation}
   \omega_{\gamma} = \frac{f_p}{\epsilon_{\gamma}}~\omega_{\gamma-{\mathrm like}} 
\end{equation}
 These results are shown in Fig.~\ref{gam_fluc_model}. 
 It is observed that the relative fluctuations of photons from the data
 are in reasonable agreement with those obtained from VENUS.
 However, the results for photons from the participant model are
 somewhat higher than those from the experimental data.

\section{TRANSVERSE ENERGY FLUCTUATIONS}

 Relativistic nuclear collisions are often described within the 
 participant-spectator picture in which nuclei are spheres that 
 collide with a definite impact parameter. The overlapping volume
 which participate in the reaction are violently disrupted while 
 the remaining spectator volumes shear off and suffer comparatively 
 mild excitations. The magnitude 
 of the $E_{\mathrm T}$ produced depends on the bombarding energy and the 
 participant volume or equivalently the number of
 participating nucleons. The cross section for a specific value of
 $E_{\mathrm T}$ production depends to a large extent 
 on the geometric probability 
 of a given impact parameter. Therefore impact parameter fluctuations
 are expected to lead to fluctuations in $E_{\mathrm T}$.

 Corroboration of the participant-spectator picture comes from the 
 strong anti-correlation of $E_{\mathrm T}$ with the energy observed in the 
 zero degree calorimeter, $E_{\mathrm F}$ as shown in the 
 Fig.~\ref{et_zdc}(b). The smaller the impact parameter, the larger 
 is the participant volume and $E_{\mathrm T}$, but the smaller is the 
 spectator volume and $E_{\mathrm F}$. $E_{\mathrm T}$ also 
 correlates strongly with the produced particle multiplicity. The 
 ${d\sigma}/{dN_{\mathrm ch}}$ and ${d\sigma}/{dN_{\gamma-{\mathrm like}}}$ 
 spectra have virtually the same shape as ${d\sigma}/{dE_{\mathrm T}}$ 
 (Fig.~\ref{nch_ngam}(a) and Fig.~\ref{nch_ngam}(b)).

 The study of the average total $E_{\mathrm T}$ as 
 measured in the WA98 experiment and its scaling behavior  with the number of
 participants have been discussed earlier in detail in Ref.~\cite{WA98-10}. 
 Here we concentrate on the second moment, and study the fluctuations in 
 $E_{\mathrm T}$ as was done for $N_{\gamma-{\mathrm like}}$ and $N_{ch}$.
 For this analysis we have again taken  $2\%$ width bins in centrality
 using the forward energy as measured by the ZDC. Due to the poorer 
 resolution in centrality 
 selection of the ZDC for peripheral collisions, we present the results 
 only up to the $50\%$ centrality class.

 The  $E_{\mathrm T}$ distribution for the top $2\%$ of the minimum bias cross
 section is shown in Fig.~{\ref{et_2perc}. The solid curve shows a Gasussian
 fit to the distribution. The $\mu$, $\sigma$ and
 ${\chi^{2}/ndf}$ values for such distributions at centrality bins
 varying from 0-2\% to 48-50\% have been extracted and
 are shown in Fig.~\ref{et_fluc_2}.  
 The ${\chi^{2}/ndf}$ values are seen
 to be between 1 and 2, which indicates that the distributions are
 well described by Gaussians. 
 The fluctuations in $E_{\mathrm T}$ have been calculated by
 using Eq.~1 and are shown in Fig.~\ref{et_omega_2}.  
 The relative 
 fluctuations are observed to increase in going from central to peripheral 
 collisions. 
 $E_{\mathrm T}$, measured by the MIRAC was used in the online trigger
 to define the most central event sample with a threshold which occurred in
 the region of the top $14\%-18\%$ of the total cross section. This
 region is not analyzed
 to avoid trigger bias effects in the measured $E_{\mathrm T}$ distribution.

 $E_{\mathrm T}$ has a strong correlation with the number of participant 
 nucleons or the number of effective
 collisions they undergo~\cite{et_baym}.  
 In an attempt to understand the fluctuations in terms of the number of 
 participant nucleons, the VENUS event generator has been used to 
 determine the ratio 
 $\mu(E_{\mathrm T})/\mu(Npart)$ as a function of centrality.  
 The $\mu(E_{\mathrm T})$ per participant has been found to be 
 $\sim 1.1 GeV \pm 0.2$. This is shown in Fig.~\ref{et_perpart}. The main
 sources of error here are due to the uncertainty in the calculation of 
 the  number of participants and the finite resolution of the the calorimeters.
 Similar results were also obtained from WA80 and HELIOS collaborations 
 Ref. \cite{wa80_et,helios}. While the WA80 Collaboration has shown
 that $E_{\mathrm T}$ per participant is independent of projectile, target,
 and centrality but depends only on the number of wounded nucleons and the 
 beam energy, the WA98 Collaboration has shown that transverse energy does
 deviate from a linear dependence on the number of participants for Pb+Pb
 collisions \cite{WA98-10}.

 The relative fluctuations in transverse energy can be analyzed in a 
 participant
 picture similar to that employed in the case of photons and charged
 particles~\cite{et_baym}. An expression similar to Eq.~4 can be obtained,
 where the first term depends on the fluctuations in the transverse
 energy deposited by each
 particle produced per participant nucleon, with the second term coming from
 impact parameter fluctuations within the acceptance of the detector. Since 
 the first term depends greatly on the detector characteristics, we compare 
 the transverse energy fluctuations in data to those obtained from a fast
 simulation of the MIRAC and ZDC characteristics in VENUS in which the energy 
 resolution for each particle was applied separately when computing the
 total transverse energy~\cite{awes}.
 Fig.~\ref{et_zdc}(a and b) shows the comparison of the simulated 
 $E_{\mathrm T}$ and $E_{\mathrm F}$ distributions 
 with those from data. The agreement is seen to be quite reasonable.  

 Fig.~\ref{et_model_2} shows the comparison of $E_{\mathrm T}$ fluctuations
 from data to those obtained from simulated events using VENUS. The 
 fluctuations
 are plotted as a function of the mean number of participants in
 various $2\%$ bins
 of centrality obtained from $E_{\mathrm F}$. Errors shown in the data
 are mainly due to uncertainties in the determination of number of 
 participants and the finite energy resolution of the calorimeters as 
 discussed earlier. It is seen that the
 fluctuation in data are systematically smaller than those obtained form
 VENUS. As discussed in Ref.~\cite{et_baym} many
 effects like energy-momentum degradation of nucleonic objects in successive
 scatterings and short range correlations between nucleons in a nucleus may
 be responsible for the decrease in fluctuations in $E_{\mathrm T}$ of data
 as compared to those obtained from simulations in VENUS. The role of
 re-scattering has also to be understood in this context.

\section{SUMMARY}

 A detailed event-by-event study of fluctuations in the multiplicities of
 charged particles and photons and transverse energy has been
 carried out using data from the WA98 experiment. This has been done 
 varying both the centrality and rapidity intervals.
 It is observed that the relative
 fluctuations increase with increase in the impact parameter interval. Hence 
 it is important to control the impact parameter dependence by studying narrow 
 bins in centrality.
 For $2\%$ centrality bins, within which the
 distributions of charged particle and photon multiplicities, as well as
 the transverse energy, are well-described by Gaussians, the
 contribution from impact 
 parameter fluctuations is around one, as expected. The fluctuations
 in multiplicities and $E_{\mathrm T}$ are found to increase in going
 from central towards peripheral events. A decrease in acceptance has
 been found to result in a decreased multiplicity fluctuations. Using a 
 simple statistical analysis one can explain the observed decrease 
 in a smaller acceptance knowing the fluctuations in a larger
 acceptance window. The observed centrality dependence of the
 multiplicity fluctuations of charged particles have been
 found to agree reasonably well with results obtained from a simple 
 participant model which takes into account impact parameter
 fluctuations from VENUS and multiplicity fluctuations from $NN$ data,
 within the quoted systematic errors. For photons the fluctuations are
 found to be slightly lower compared to those obtained from the participant 
 model. 

 Similar calculations have been performed for the transverse energy and 
 multiplicity distributions using
 VENUS. The transverse energy fluctuations from experimental data 
 are found to be 
 smaller than those observed in VENUS. 
 On the other hand, after corrections for charged particle 
 contamination in the
 photon-like clusters, the relative fluctuations of photons appear to be
 in rather good agreement with VENUS.

\medskip

{\bf Acknowledgements} \\

We wish to express our gratitude to the CERN accelerator division for the
excellent performance of the SPS accelerator complex. We acknowledge with
appreciation the effort of all engineers, technicians and support staff who
have participated in the construction of this experiment.

This work was supported jointly by
the German BMBF and DFG,
the U.S. DOE,
the Swedish NFR and FRN,
the Dutch Stichting FOM,
the Polish KBN under Contract No. 621/E-78/SPUB-M/CERN/P-03/DZ211/,
the Grant Agency of the Czech Republic under contract No. 202/95/0217,
the Department of Atomic Energy,
the Department of Science and Technology,
the Council of Scientific and Industrial Research and
the University Grants
Commission of the Government of India,
the Indo-FRG Exchange Program,
the PPE division of CERN,
the Swiss National Fund,
the INTAS under Contract INTAS-97-0158,
ORISE,
Grant-in-Aid for Scientific Research
(Specially Promoted Research \& International Scientific Research)
of the Ministry of Education, Science and Culture,
the University of Tsukuba Special Research Projects, and
the JSPS Research Fellowships for Young Scientists.
ORNL is managed by UT-Battelle, LLC, for the U.S. Department of Energy
under contract DE-AC05-00OR22725.
The MIT group has been supported by the US Dept. of Energy under the
cooperative agreement DE-FC02-94ER40818.

\newpage

\onecolumn

\begin{figure}
\begin{center}
\includegraphics[scale=0.4]{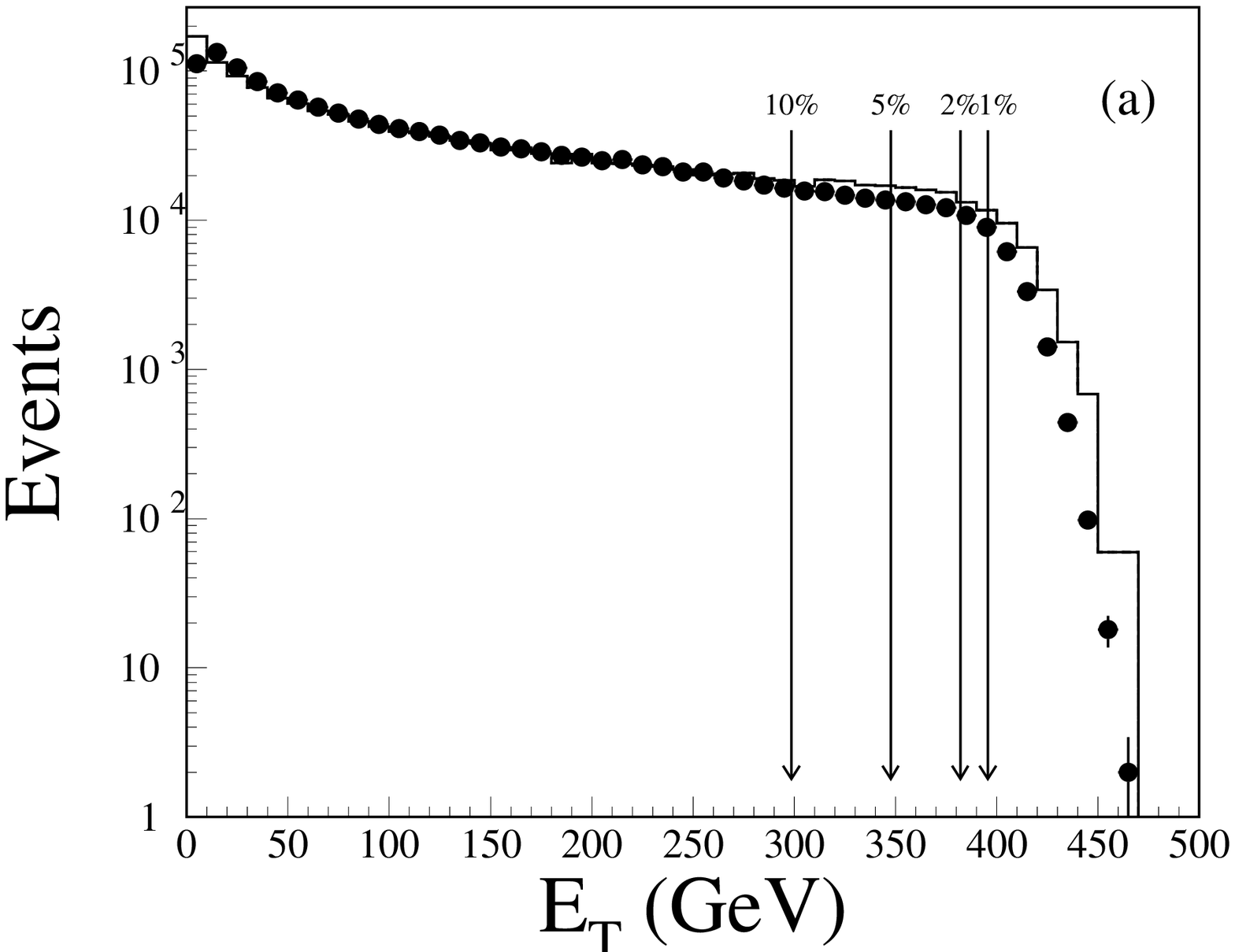}
\includegraphics[scale=0.4]{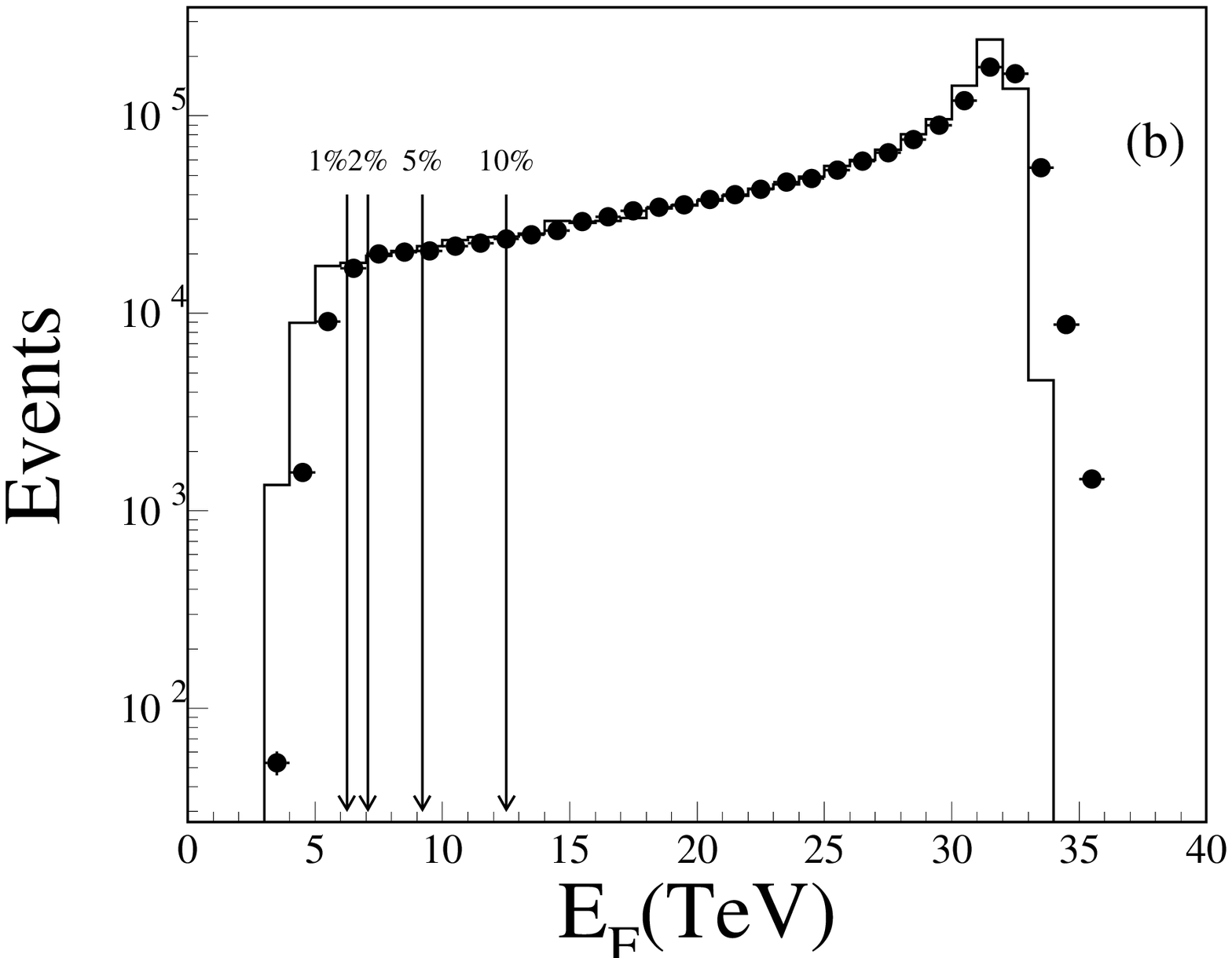}
\includegraphics[scale=0.4]{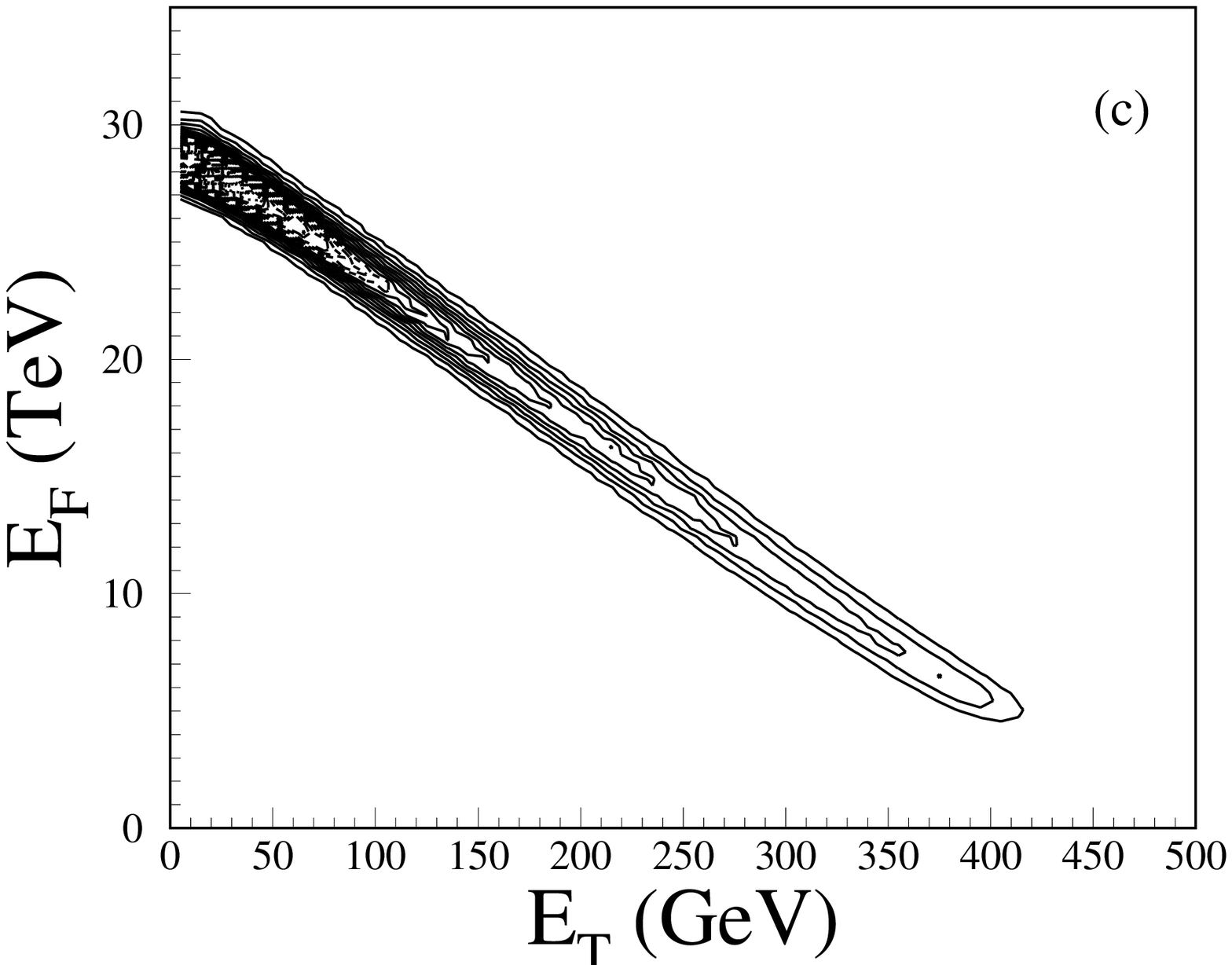}
\caption{ Minimum bias distributions of (a) transverse energy $E_{\mathrm T}$,
(b) forward energy $E_{\mathrm F}$ produced 
    in Pb induced reactions at 158$\cdot A$ GeV on Pb. 
    Solid histograms show the results obtained from VENUS event generator.
(c) shows the anti-correlation of measured total transverse energy,
    $E_{\mathrm T}$, and forward energy, $E_{\mathrm F}$.
}
\label{et_zdc}
\end{center}
\end{figure}

\begin{figure}
\begin{center}
\includegraphics[scale=0.4]{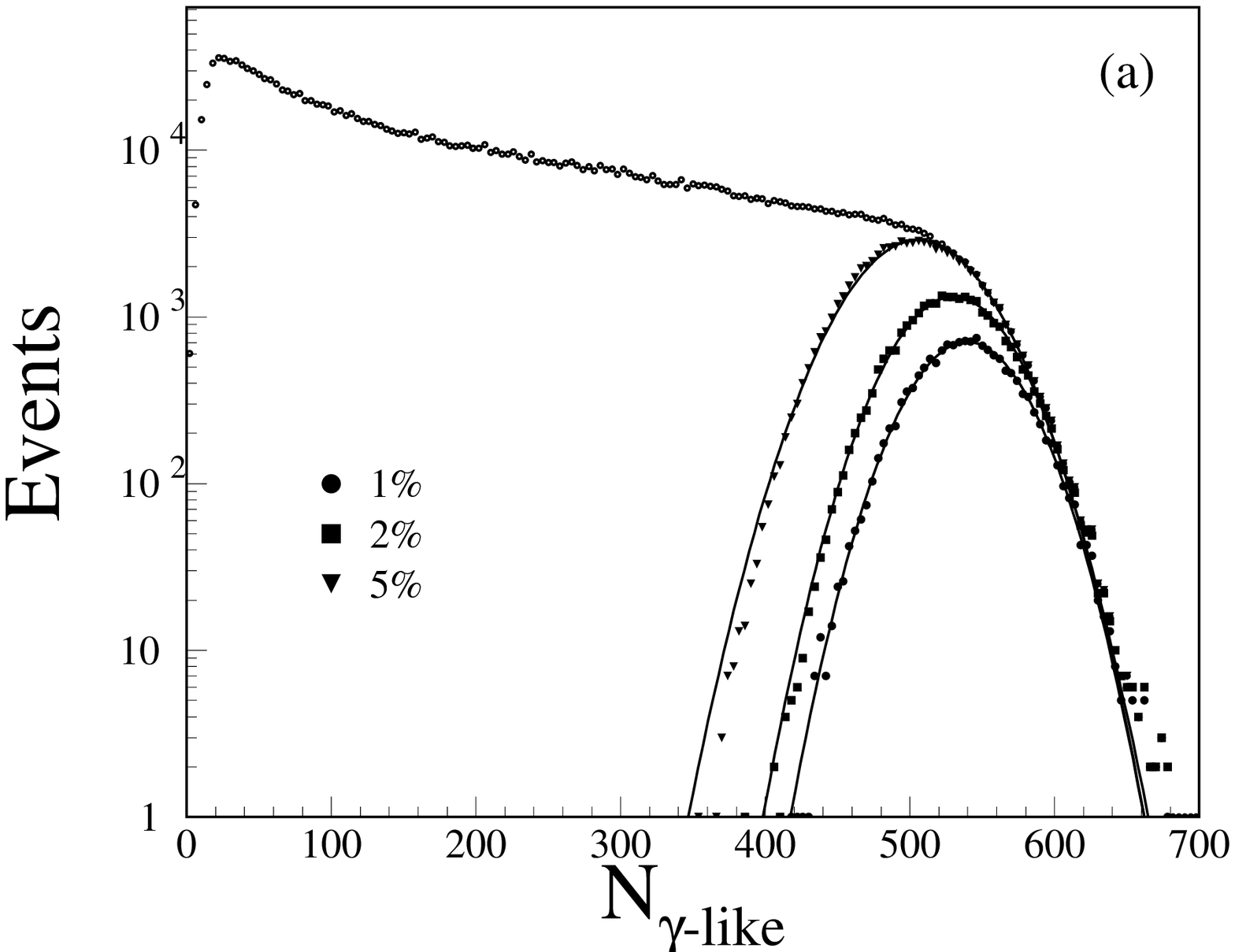}
\includegraphics[scale=0.4]{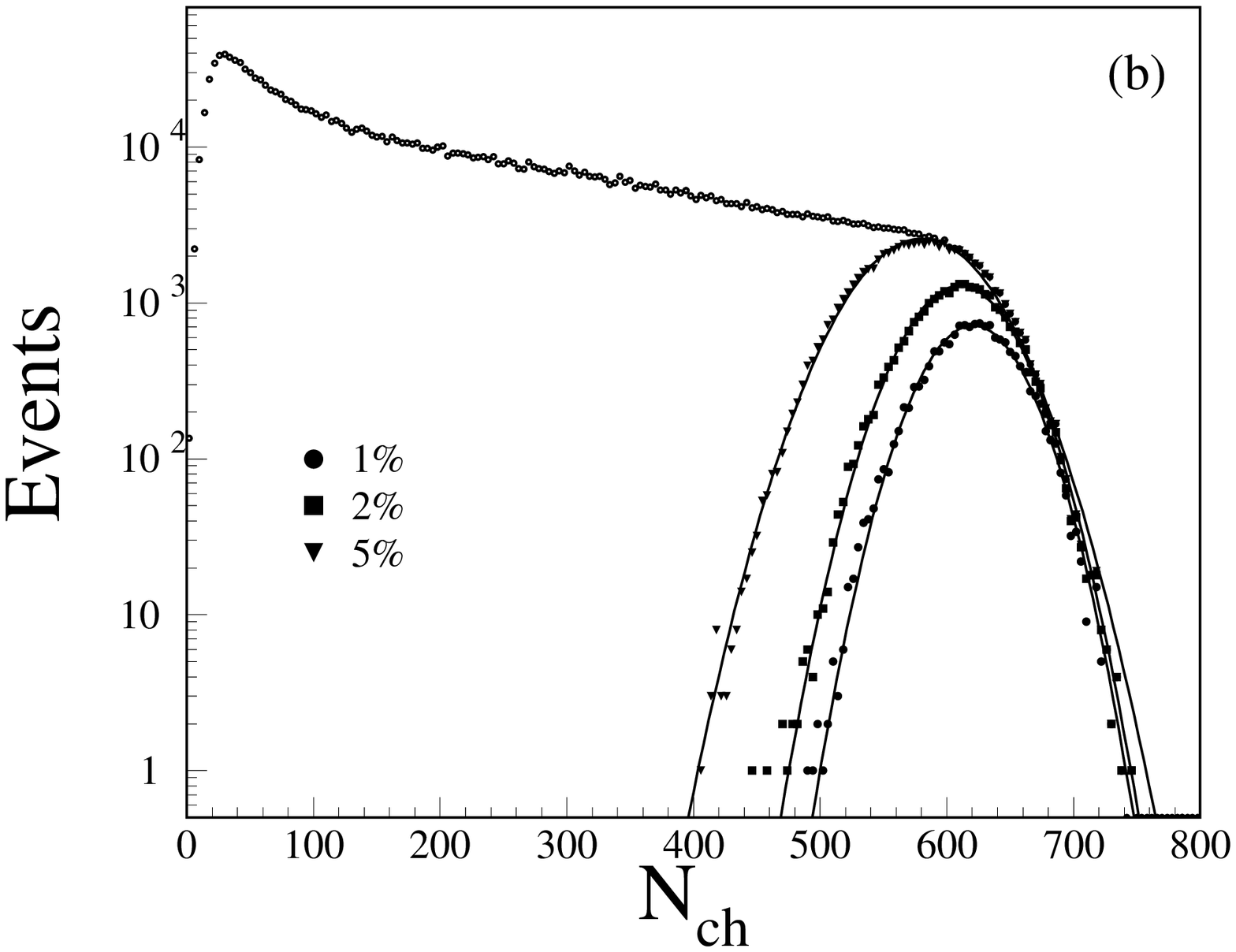}
\caption{ Minimum bias distributions of 
(a) $\gamma$-like cluster multiplicity, and
(b) charged particle multiplicity produced 
    in Pb induced reactions at 158$\cdot A$ GeV on Pb.
The multiplicity distributions for the top $1\%$, $2\%$, and $5\%$
most central events are also shown and fitted to 
Gaussian distributions.
}
\label{nch_ngam}
\end{center}
\end{figure}

\begin{figure}
\begin{center}
\includegraphics[scale=0.4]{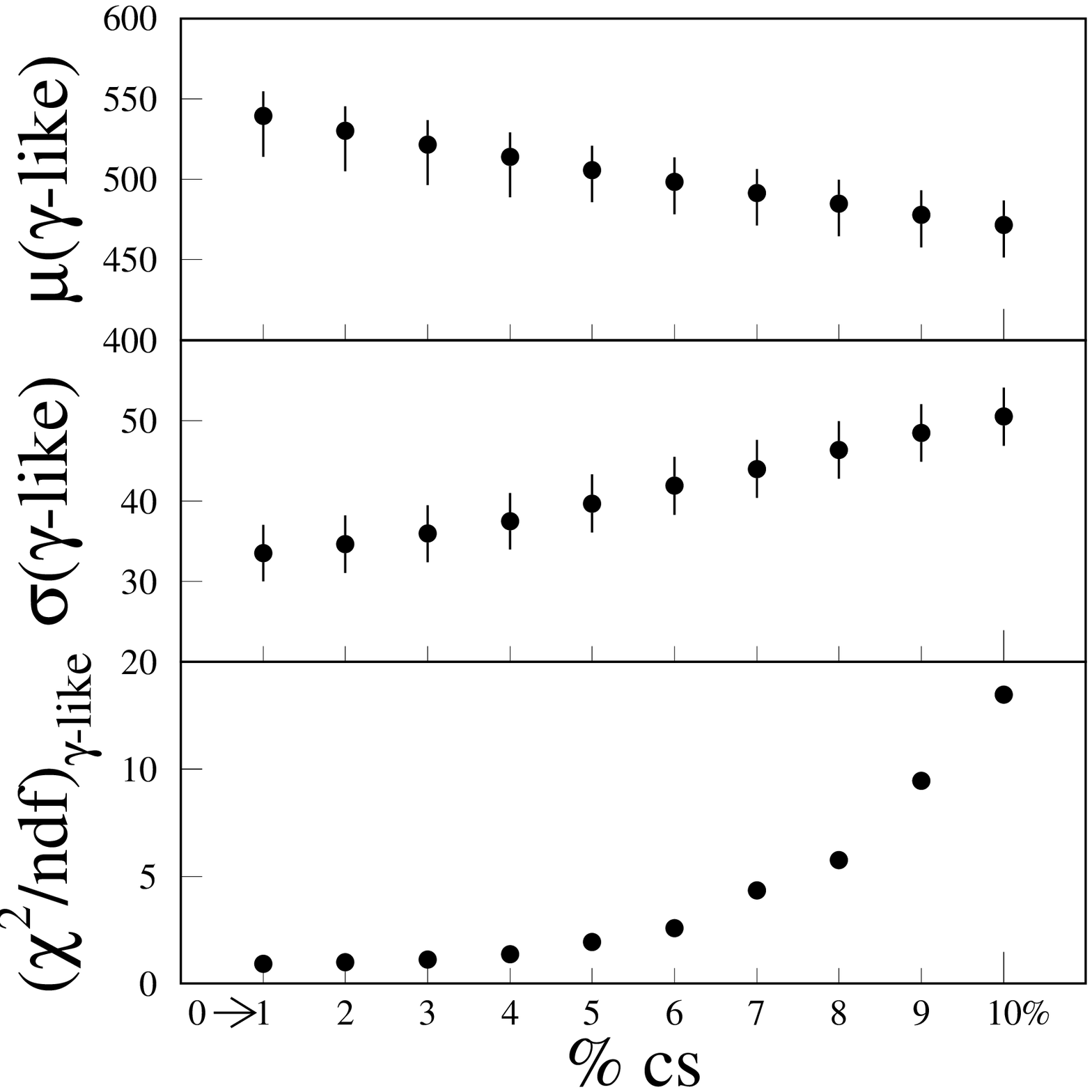}
\includegraphics[scale=0.4]{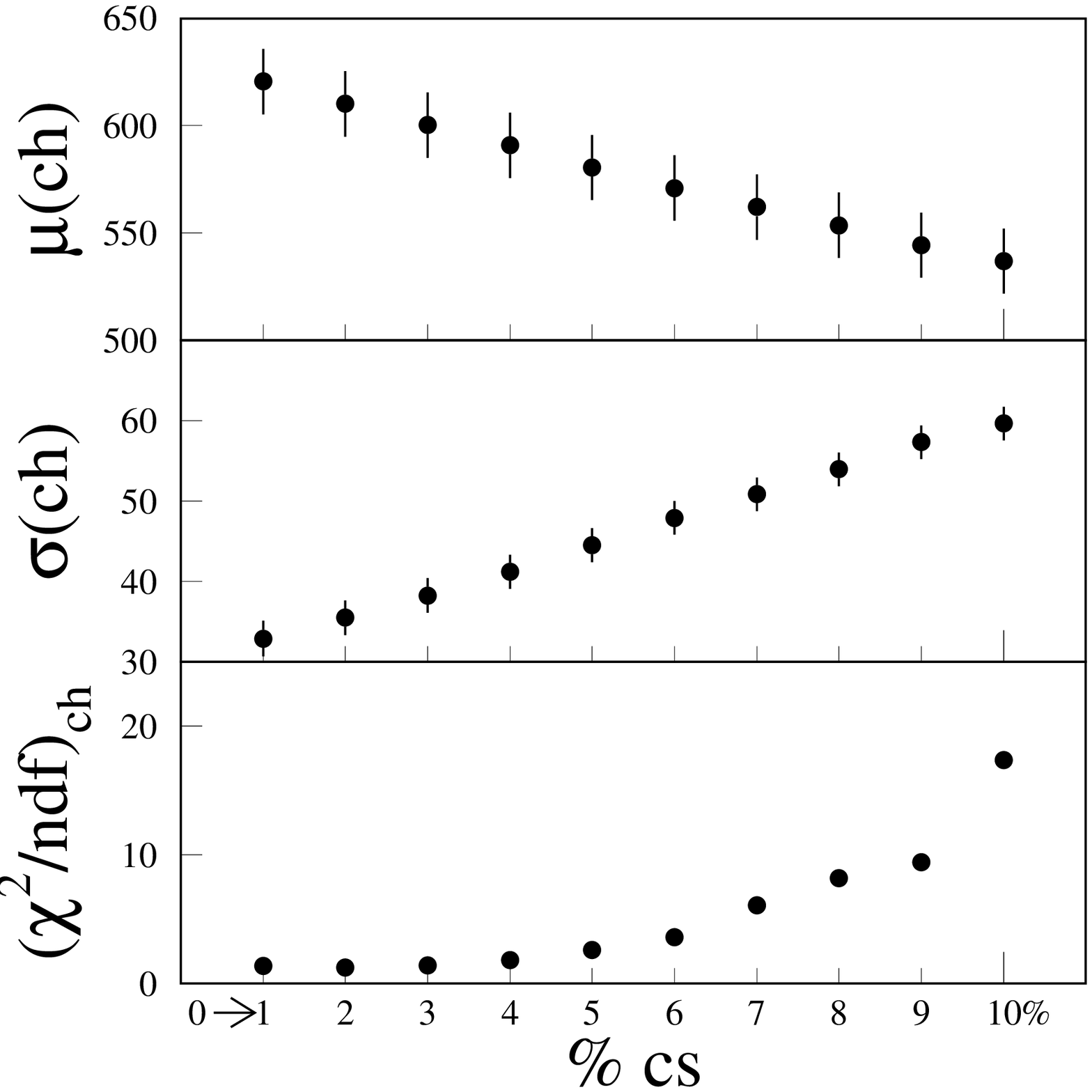}
\bigskip
\caption{ 
Gaussian fit parameters of the multiplicity distributions of
$\gamma$-like clusters and charged particles for increasing
centrality bins of increasing width. 
The centrality selection has been made by increasing the widths
of $E_{\mathrm T}$ bins corresponding to
$0-1\%$, $0-2\%$, $0-3\%$,$\cdots$,$0-10\%$ of the minimum bias cross section.
}
\label{cen_nch_ngam}
\end{center}
\end{figure}

\begin{figure}
\begin{center}
\includegraphics[scale=0.4]{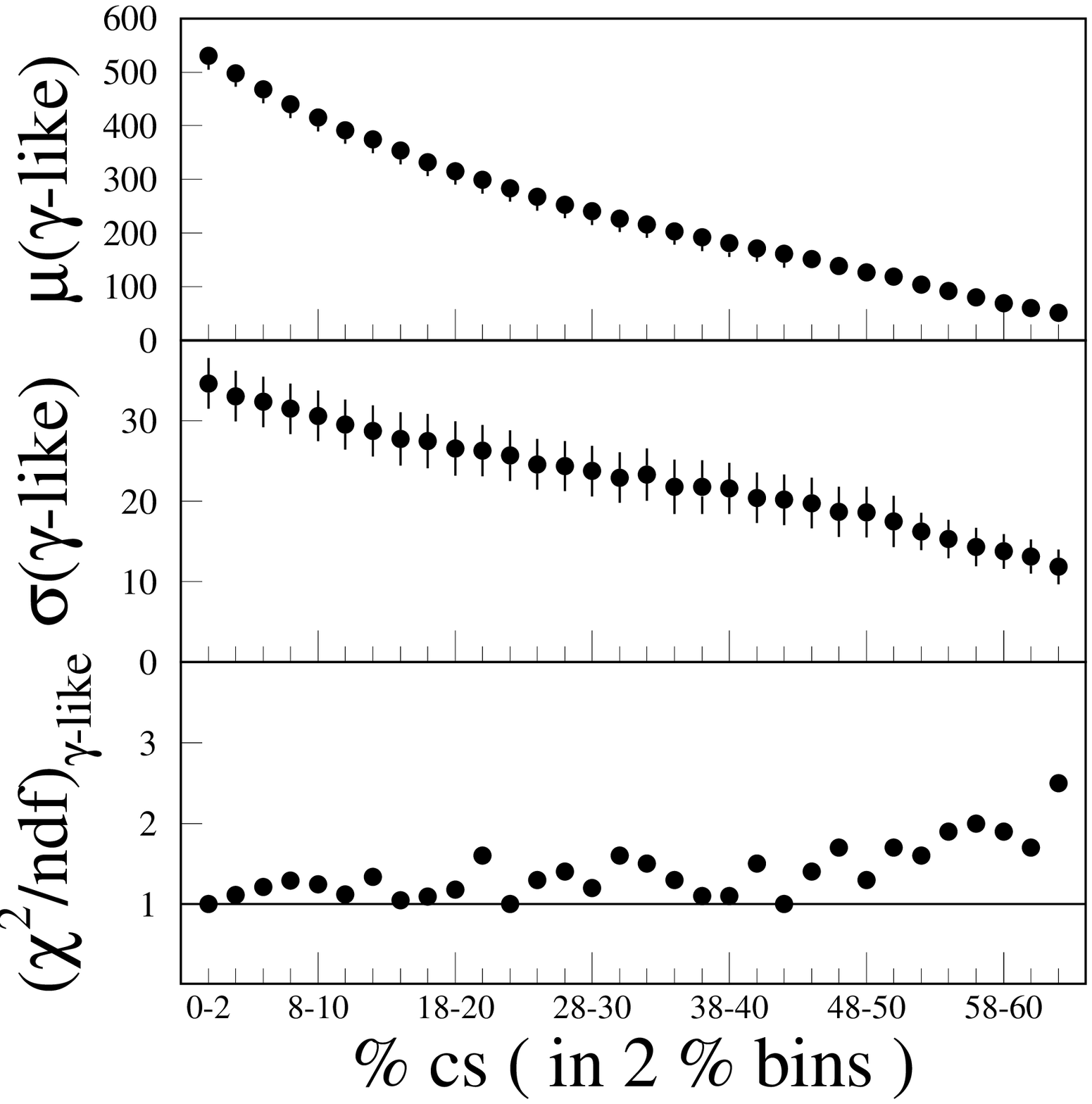}
\includegraphics[scale=0.4]{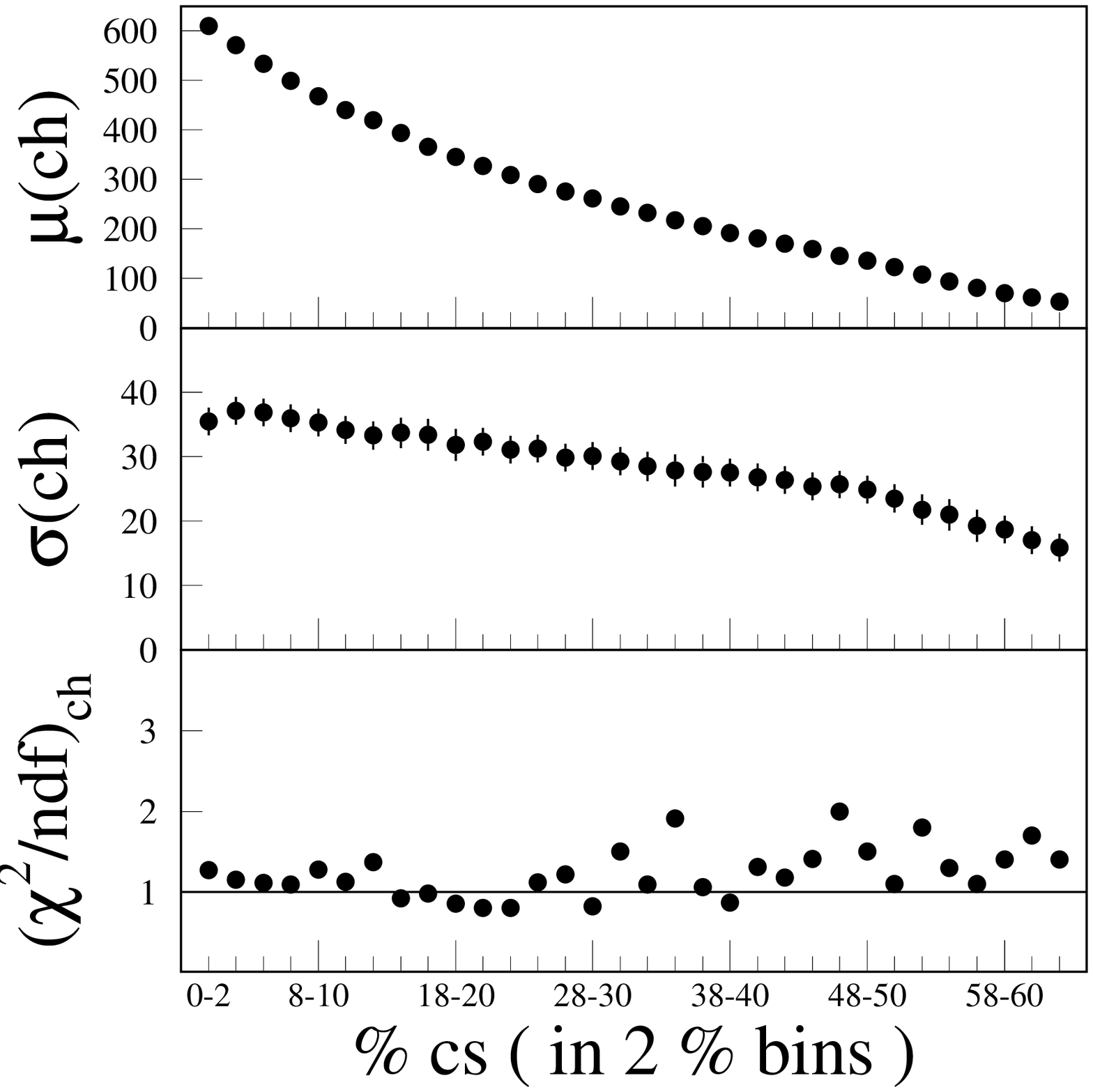}
\bigskip
\caption{ 
Gaussian fit parameters of the multiplicity distributions of
$\gamma-like$ clusters and charged particles as a function of
centrality. The centrality selection has been made by selecting
$2\%$ bins in minimum bias cross section, 
$viz.$, $0-2\%$, $2-4\%$, $4-6\%$,$\cdots$,$62-64\%$. The multiplicity
distributions for these centrality bins are near perfect Gaussians as
can be seen from the $\chi^2/ndf$ values.
}
\label{cen2_nch_ngam}
\end{center}
\end{figure}

\begin{figure}
\begin{center}
\includegraphics[scale=0.4]{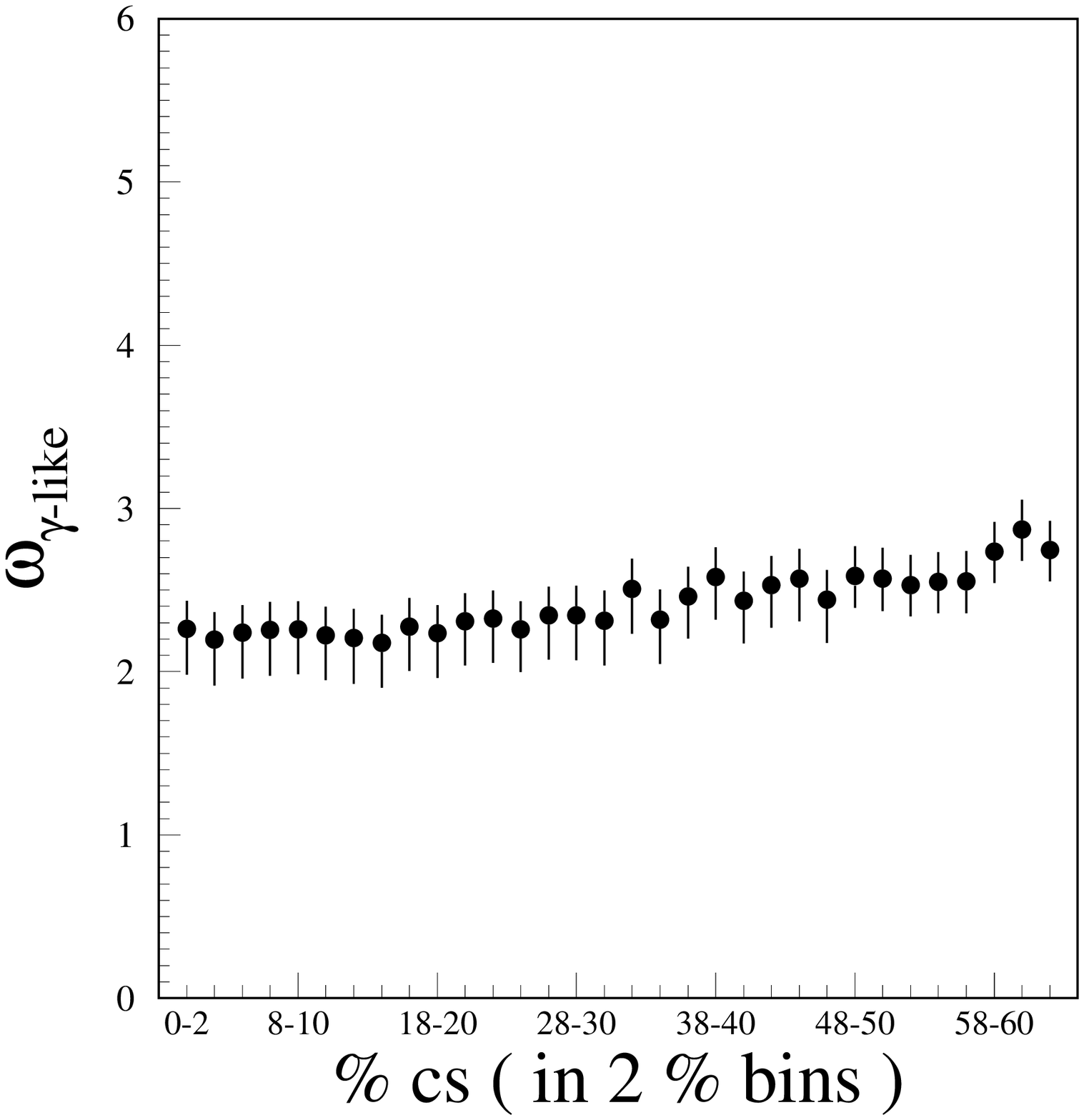}
\includegraphics[scale=0.4]{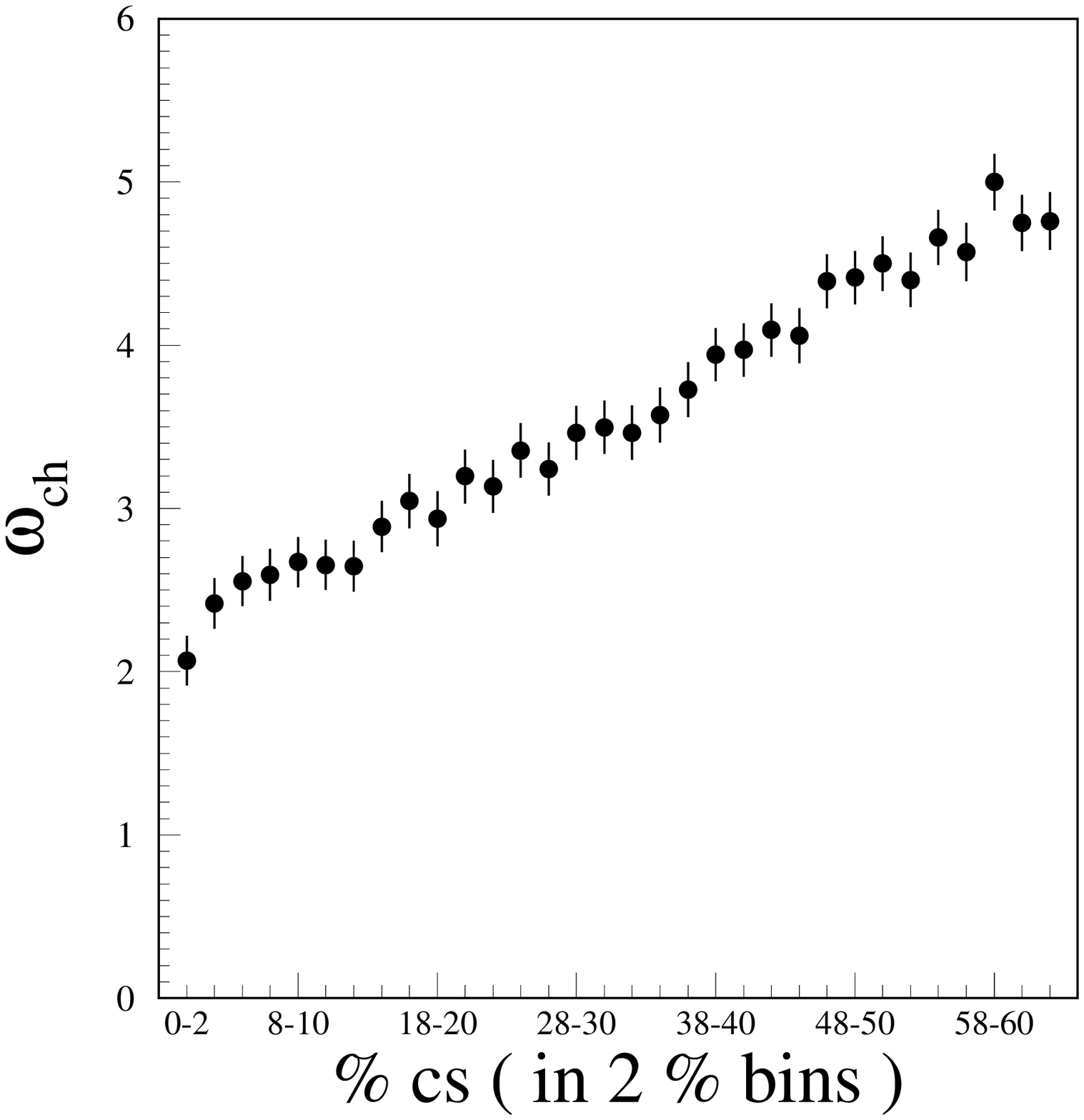}
\bigskip
\caption{ 
Fluctuations of the multiplicity of 
$\gamma$-like clusters and charged particles within the full coverage 
of PMD and SPMD
for various $2\%$ bins of the minimum bias cross section.
}
\label{2perc_W}
\end{center}
\end{figure}

\begin{figure}
\begin{center}
\includegraphics[scale=0.4]{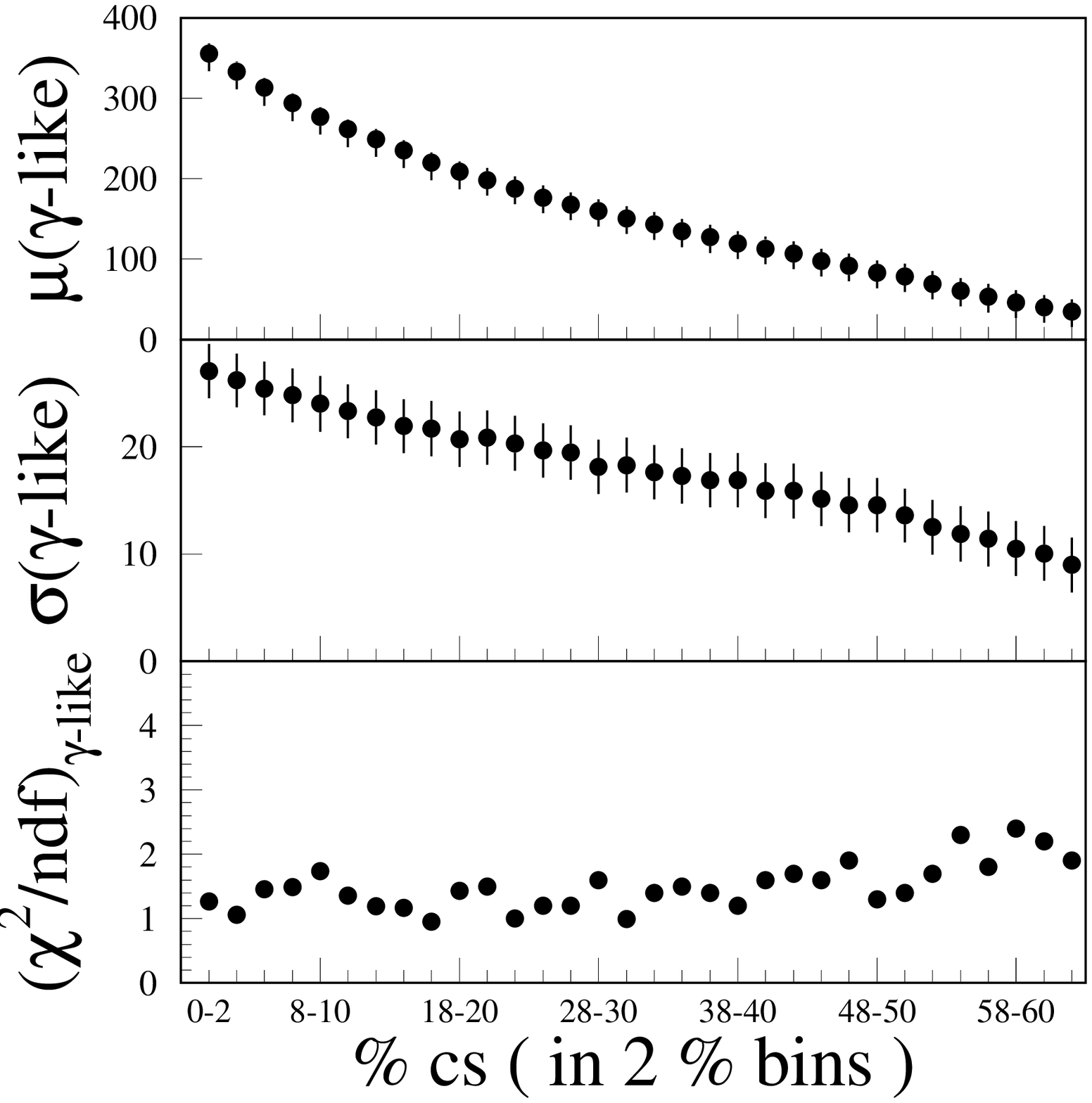}
\includegraphics[scale=0.4]{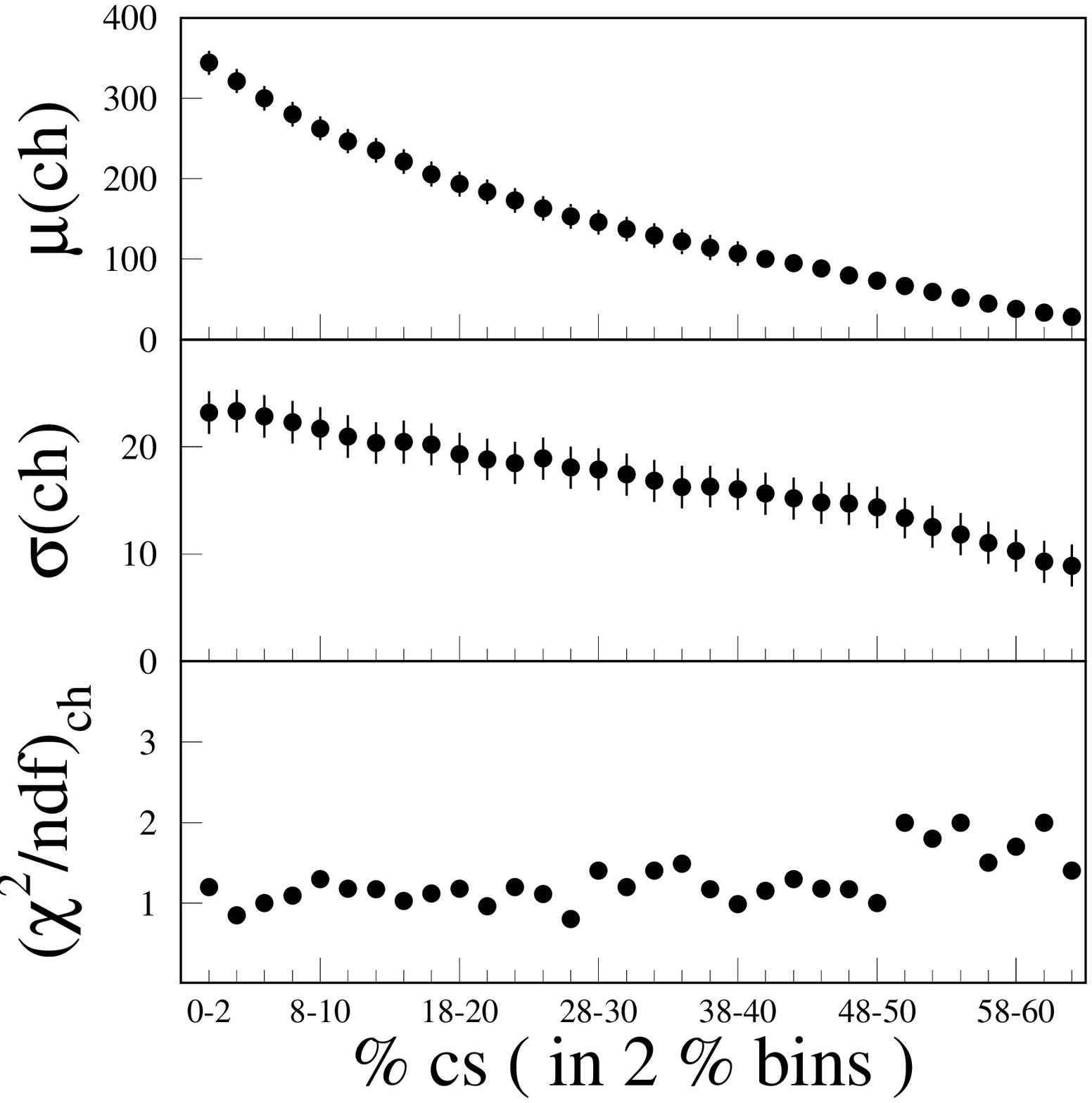}
\bigskip
\caption {
Centrality dependence of the 
Gaussian fit parameters of the multiplicity distribution of $\gamma$-like 
clusters and charged
particles within the common coverage of PMD and SPMD.
}
\label{com_ngam_nch}
\end{center}
\end{figure}

\begin{figure}
\begin{center}
\includegraphics[scale=0.4]{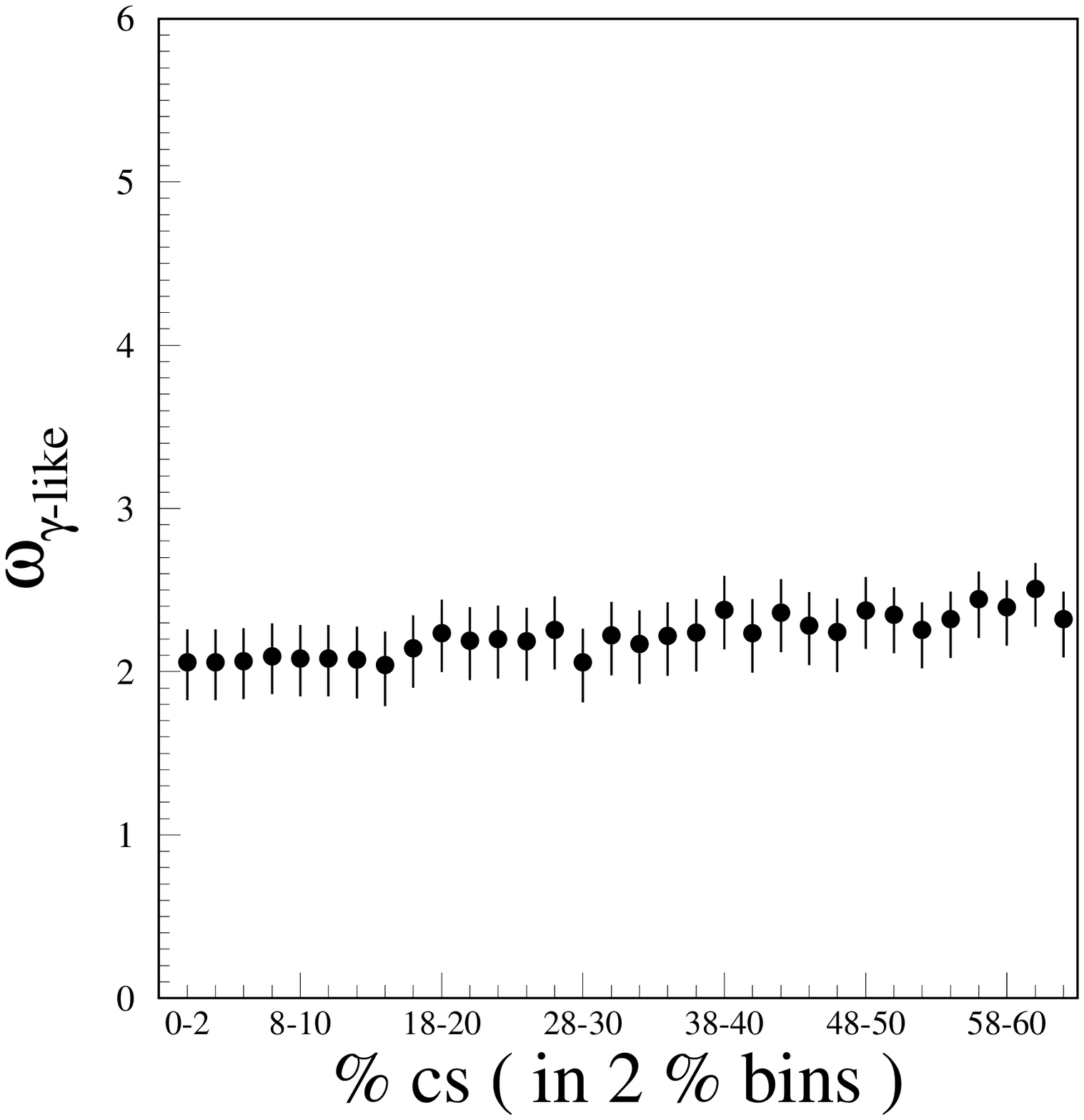}
\includegraphics[scale=0.4]{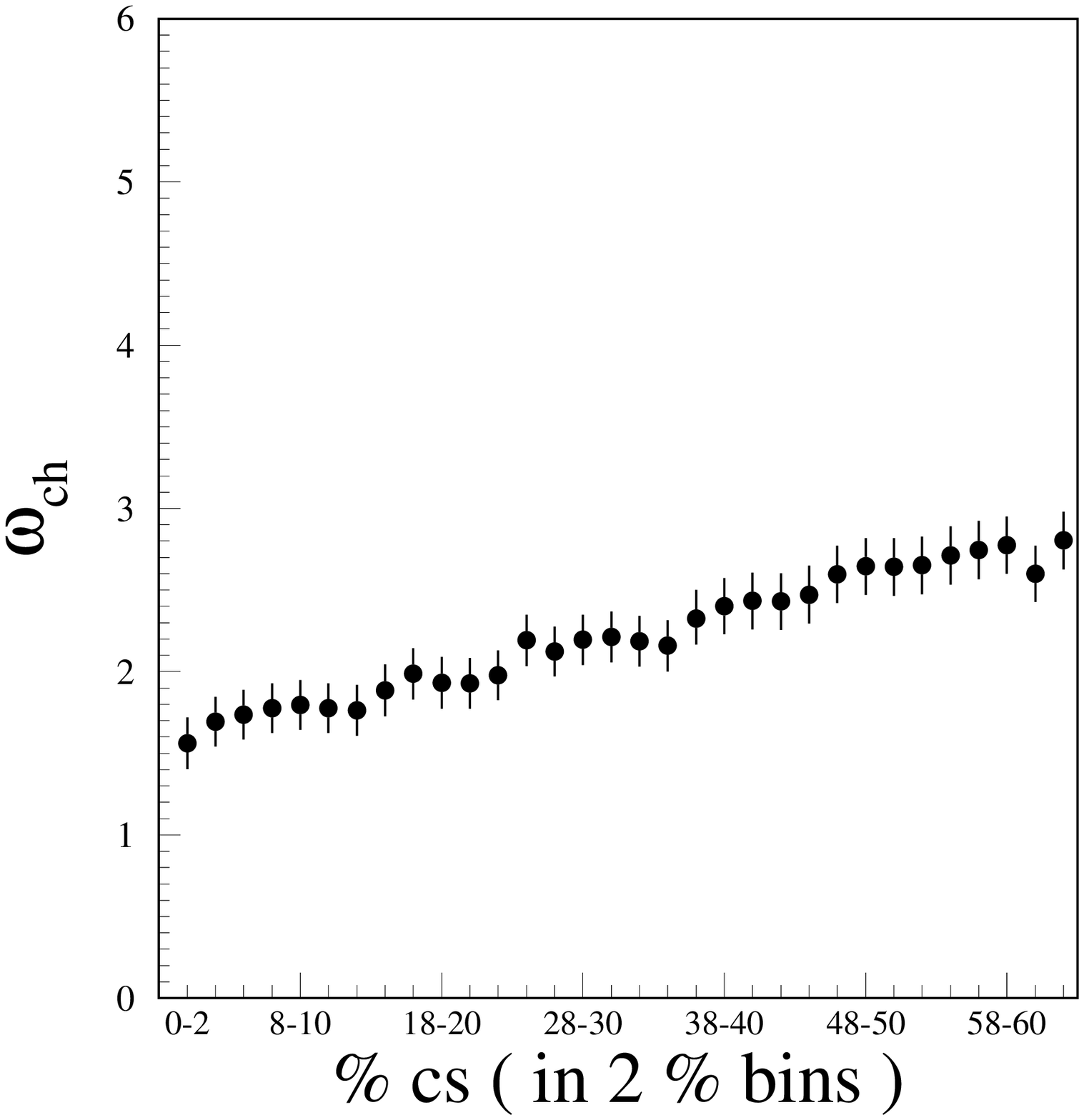}
\bigskip
\caption{ 
Centrality dependence of the fluctuations of the multiplicity
of photons and charged particles  
within the common coverage of the PMD and SPMD.
}
\label{com_fluc}
\end{center}
\end{figure}

\begin{figure}
\begin{center}
\includegraphics[scale=0.4]{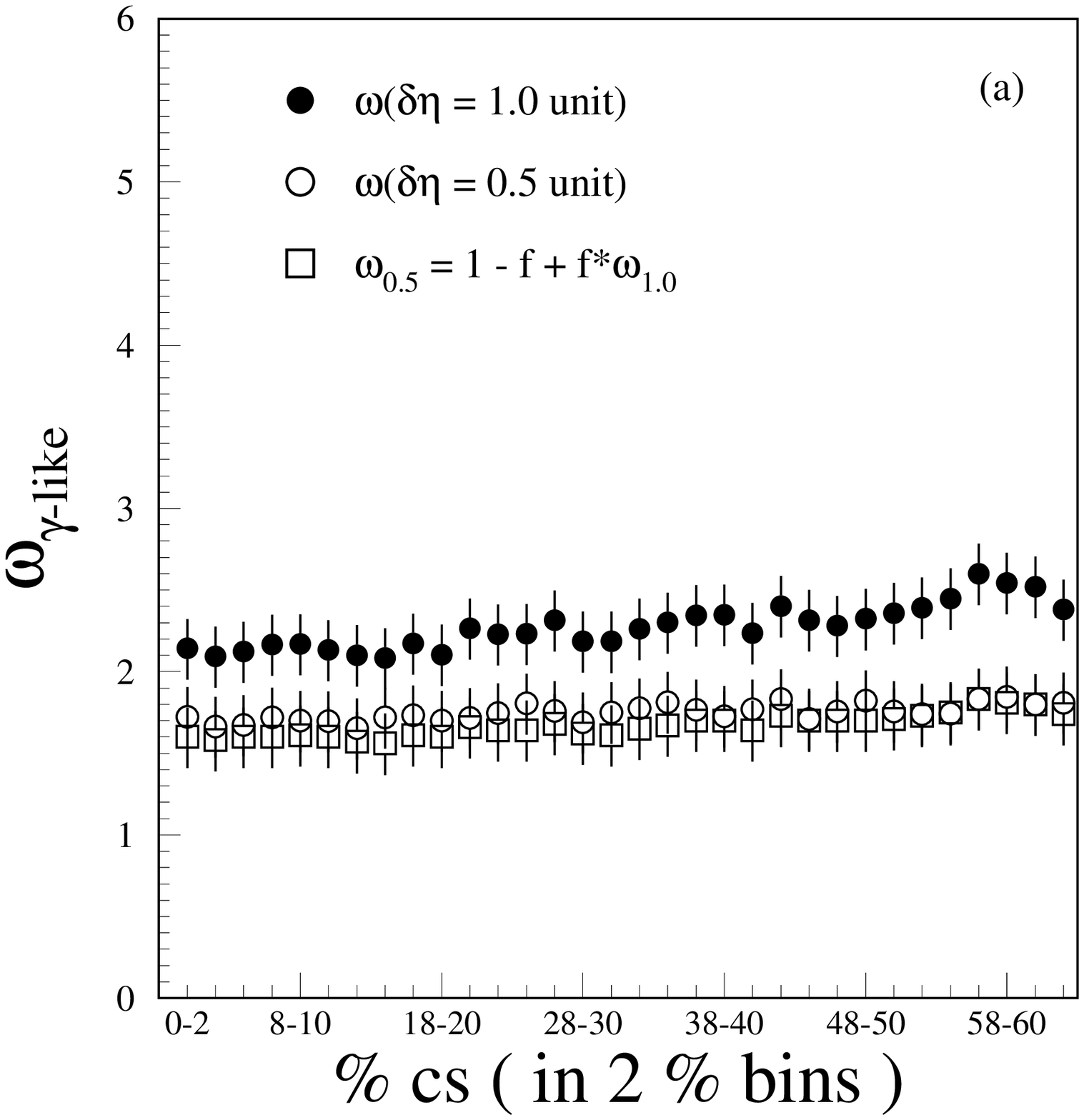}
\includegraphics[scale=0.4]{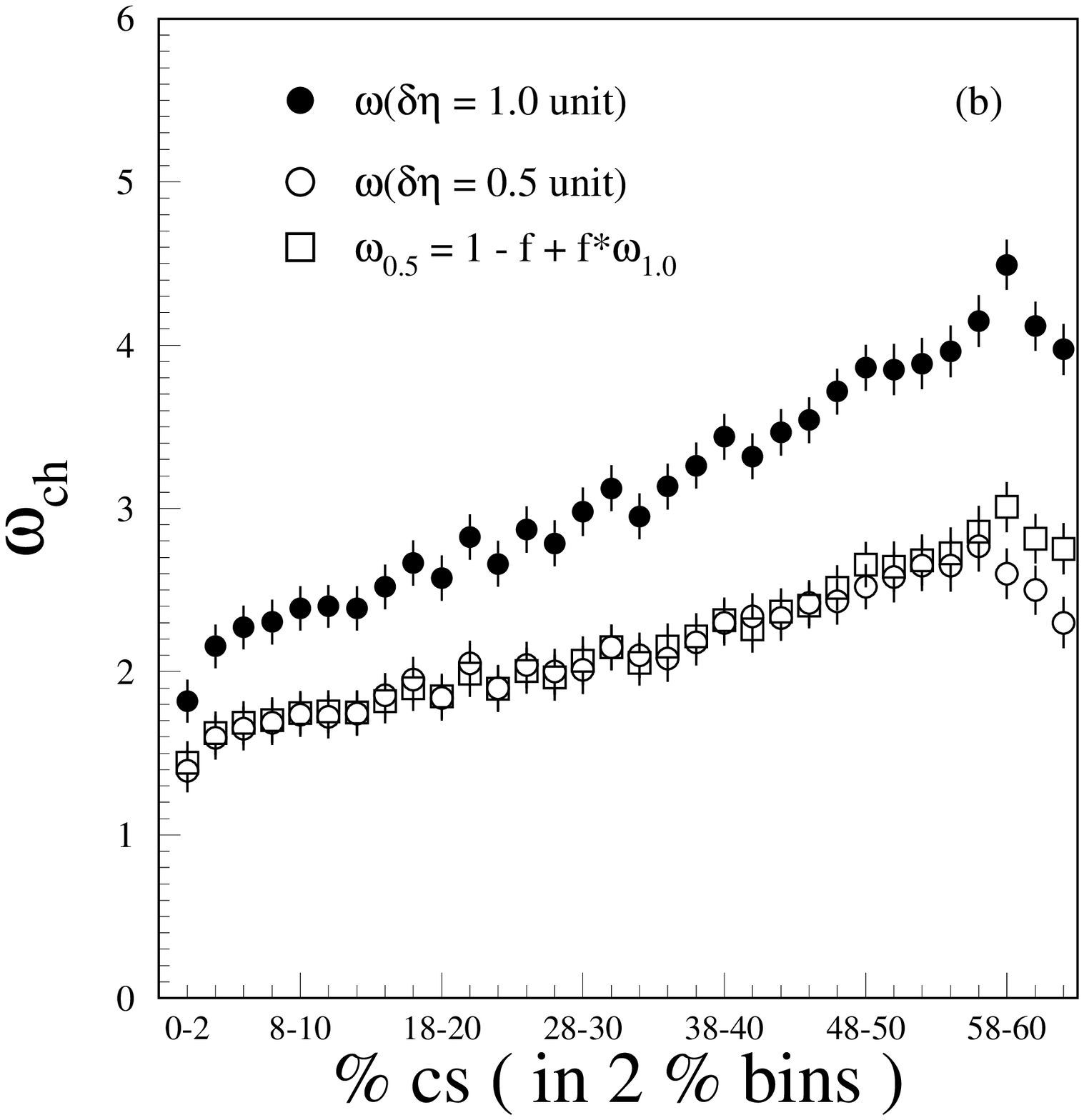}
\bigskip
\caption {\label{fluc_acc}
Multiplicity fluctuations 
of photons and charged particles for two $\eta$ acceptance selections.
The open squares represent estimated values fluctuations
in $0.5$ unit of $\delta \eta$ from the observed 
fluctuations in $1.0$ unit of $\delta \eta$.
}
\end{center}
\end{figure}

\newpage

\begin{figure}
\begin{center}
\includegraphics[scale=0.4]{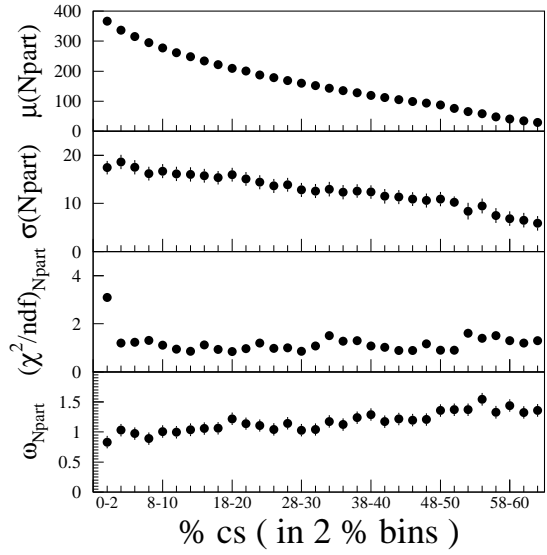}
\bigskip
\caption {\label{npart_fluc_2}
Variation of $\mu$, $\sigma$, and ${\chi^{2}/ndf}$ of the 
distribution of the number 
of participants as a function of centrality.
}
\end{center}
\end{figure}

\begin{figure}
\begin{center}

\includegraphics[scale=0.4]{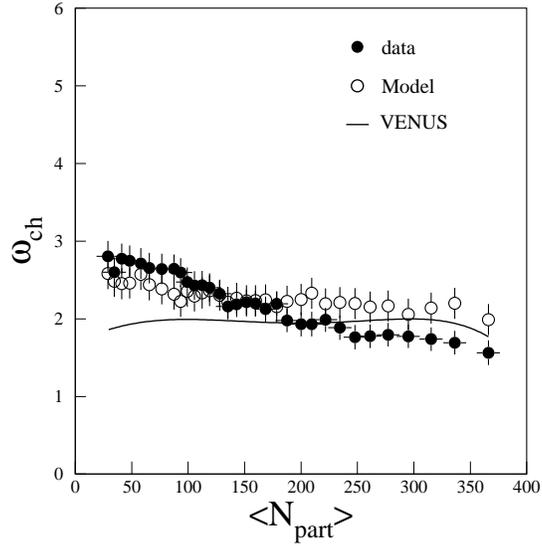}

\bigskip
\caption {\label{ch_fluc_model}
The relative fluctuations, $\omega_{ch}$, of the charged particle 
multiplicity as a function of number of participants. The experimental data
are compared to 
calculations from a participant model and those from VENUS event generator.
}
\end{center}
\end{figure}

\begin{figure}
\begin{center}
\includegraphics[scale=0.4]{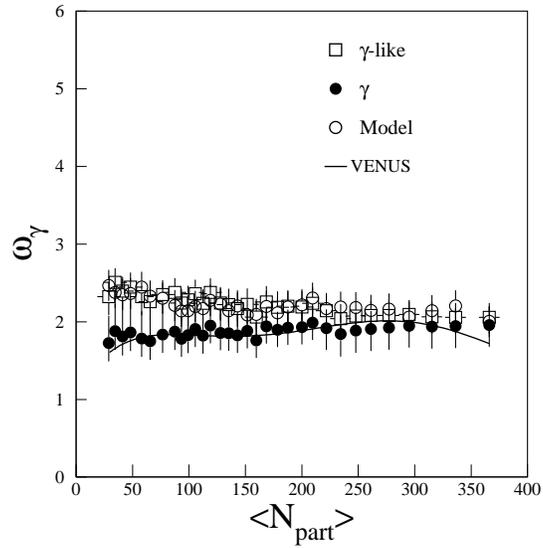}
\bigskip
\caption {\label{gam_fluc_model}
The relative fluctuations, $\omega_{\gamma}$ of photons
as a function of number of participants.
The data presented show the
fluctuations in $\gamma-like$ clusters and photons
after correction for efficiency and
purity. These are compared to 
calculations from a participant model and those from VENUS event generator.
}
\end{center}
\end{figure}

\begin{figure}
\begin{center}
\includegraphics[scale=0.4]{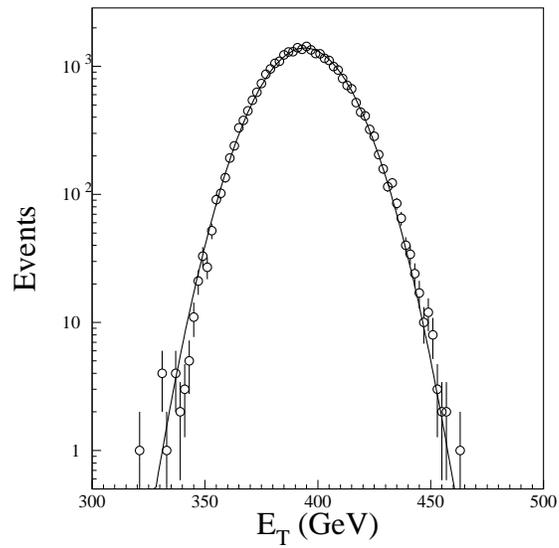}
\bigskip
\caption {\label{et_2perc}
The transverse energy distribution for the top 2\% of the 
minimum bias cross section.
}
\end{center}
\end{figure}

\begin{figure}
\begin{center}
\epsfig{figure=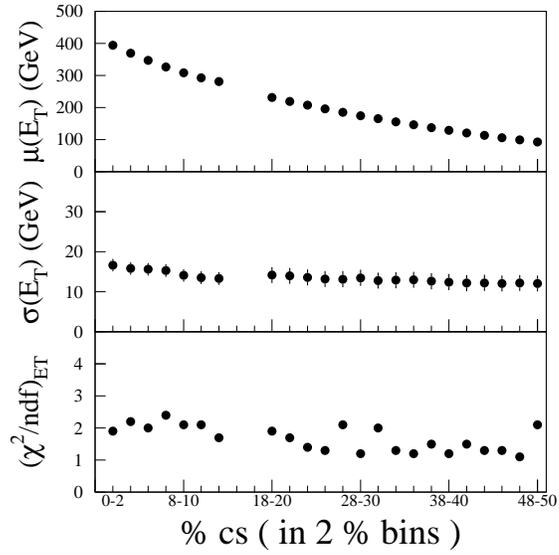,width=8.0cm}
\bigskip
\caption {\label{et_fluc_2}
Centrality dependence of $\mu$, $\sigma$, and ${\chi^{2}/ndf}$ of 
the transverse energy distribution. The centrality selection 
is base on $E_{\mathrm F}$ measured with the  ZDC. 
}
\end{center}
\end{figure}

\begin{figure}
\begin{center}
\epsfig{figure=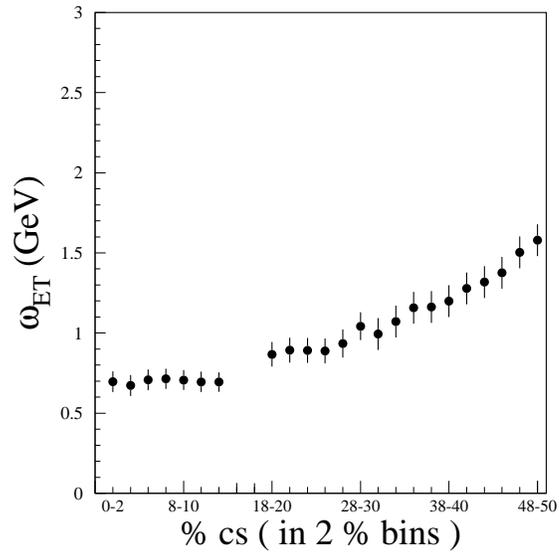,width=8.0cm}
\bigskip
\caption {\label{et_omega_2}
Centrality dependence of the relative 
fluctuations in total transverse energy, $E_{\mathrm T}$, 
with centrality selected by $E_{\mathrm F}$.
}
\end{center}
\end{figure}

\begin{figure}
\begin{center}
\epsfig{figure=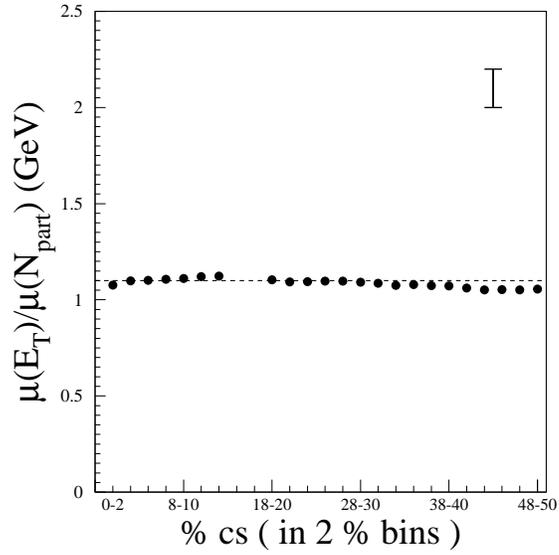,width=8.0cm}
\bigskip
\caption {\label{et_perpart}
$E_{\mathrm T}$ per participant as a function of centrality.
The vertical solid line indicates the estimated systematic
error in $E_{\mathrm T}$ per participant.
}
\end{center}
\end{figure}

\begin{figure}
\begin{center}
\epsfig{figure=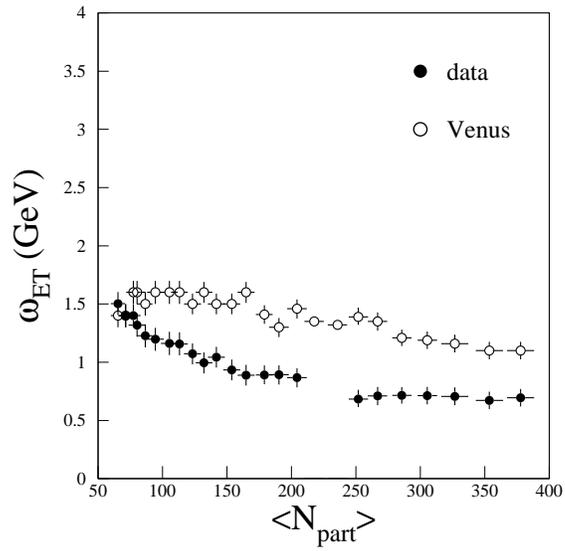,width=8.0cm}
\bigskip
\caption {\label{et_model_2}
Centrality dependence of the 
relative fluctuations in transverse energy, $E_{\mathrm T}$,
with the centrality selected from the $E_{\mathrm F}$. The result
is compared to VENUS using similar centrality selection criteria.
}
\end{center}
\end{figure}

\end{document}